\PassOptionsToPackage{table}{xcolor}
\documentclass[acmsmall, screen]{acmart}
\microtypesetup{expansion=false}
\AtBeginDocument{%
  }

\setcopyright{none}
\renewcommand{\footnotetextcopyrightpermission}[1]{}
\newcommand{\myparagraph}[1]{\noindent{\bf {#1}.}}

\usepackage[breakable,skins]{tcolorbox}
\newcommand{\takeaway}[1]{%
    \begin{tcolorbox}[
        breakable,
        colback=gray!10,
        colframe=gray!45!black,
        boxrule=0.6pt,
        arc=2pt,
        left=0.8mm, right=0.8mm, top=0.6mm, bottom=0.6mm,
        before skip=4pt, after skip=4pt,
    ]%
    #1
    \end{tcolorbox}%
}
\newcommand{\rqbox}[2]{%
    \begin{tcolorbox}[
        breakable,
        colback=#1!15,
        colframe=#1!50!black,
        boxrule=0.5pt,
        arc=2pt,
        left=0.8mm, right=0.8mm, top=0.6mm, bottom=0.6mm,
        before skip=3pt, after skip=3pt,
    ]%
    #2%
    \end{tcolorbox}%
}
\newcommand{\projectname}{Oraqle}

\usepackage{tikz}
\usepackage{hyperref}
\usepackage{quantikz}
\usepackage{makecell}
\graphicspath{ {./figures/} }
\usetikzlibrary{%
    shapes,
    arrows,
    arrows.meta,
    decorations.pathreplacing,
    calc,
}
\usepackage{subcaption}
\usepackage{multirow}
\usepackage{float}
\usepackage{pifont}     
\usepackage{array}      
\usepackage{xcolor}
\usepackage{wrapfig}
\usepackage{booktabs}
\usepackage{longtable}
\usepackage{colortbl}
\usepackage{arydshln} 
\usepackage{hhline}
\usepackage{grumble}
\usepackage{placeins}
\usepackage{nicematrix}

\usepackage{tcolorbox}
\tcbuselibrary{skins}

\usepackage[inline]{enumitem}
\setlist{noitemsep,topsep=0pt,parsep=0pt,partopsep=0pt}

\usepackage[subtle]{savetrees}

\usepackage{setspace}
\usepackage[margin=5pt,font={stretch=0.9}]{caption}

\newcommand{\herqulesnote}{For consistency with the other models, we evaluate \textsc{HERQULES} using a deployment-faithful \emph{per-trace} protocol (one shot, one prediction) and retrain it at every duration; see App.~\ref{app:herqules-bug} for details.}

\begin{document}

\title{\projectname: An Empirical Analysis of Qubit Readout and Discriminators in Quantum Error Correction}

\definecolor{takeawayblue}{HTML}{f0f4fa}
\definecolor{takeawayborder}{HTML}{4a7ab5}
\definecolor{takeawayorangebg}{HTML}{fbf1e4}
\definecolor{takeawayorangeborder}{HTML}{d79b00}
\definecolor{takeawaygreenbg}{HTML}{eaf3e7}
\definecolor{takeawaygreenborder}{HTML}{4a9b4a}
\definecolor{takeawaypurplebg}{HTML}{f2ecf8}
\definecolor{takeawaypurpleborder}{HTML}{7a52a8}
\newcommand{\takeawayblue}[1]{%
  \begin{tcolorbox}[enhanced,colback=takeawayblue,colframe=takeawayblue,boxrule=0pt,arc=0pt,left=6pt,right=4pt,top=2pt,bottom=2pt,boxsep=0pt,before skip=0pt,after skip=0pt,borderline west={2pt}{0pt}{takeawayborder}]
  #1
  \end{tcolorbox}%
}
\newcommand{\takeawayorange}[1]{%
  \begin{tcolorbox}[enhanced,colback=takeawayorangebg,colframe=takeawayorangebg,boxrule=0pt,arc=0pt,left=6pt,right=4pt,top=2pt,bottom=2pt,boxsep=0pt,before skip=0pt,after skip=0pt,borderline west={2pt}{0pt}{takeawayorangeborder}]
  #1
  \end{tcolorbox}%
}
\newcommand{\takeawaygreen}[1]{%
  \begin{tcolorbox}[enhanced,colback=takeawaygreenbg,colframe=takeawaygreenbg,boxrule=0pt,arc=0pt,left=6pt,right=4pt,top=2pt,bottom=2pt,boxsep=0pt,before skip=0pt,after skip=0pt,borderline west={2pt}{0pt}{takeawaygreenborder}]
  #1
  \end{tcolorbox}%
}
\newcommand{\takeawaypurple}[1]{%
  \begin{tcolorbox}[enhanced,colback=takeawaypurplebg,colframe=takeawaypurplebg,boxrule=0pt,arc=0pt,left=6pt,right=4pt,top=2pt,bottom=2pt,boxsep=0pt,before skip=0pt,after skip=0pt,borderline west={2pt}{0pt}{takeawaypurpleborder}]
  #1
  \end{tcolorbox}%
}


\author{Emmanouil Giortamis}
\email{emmanouil.giortamis@tum.de}
\orcid{0009-0000-3638-2969}
\affiliation{%
  \institution{Technical University of Munich}
  \country{Germany}
}

\author{Aleksandra Świerkowska}
\email{aleksandra.swierkowska@tum.de}
\orcid{0009-0008-8688-0155}
\affiliation{%
  \institution{Technical University of Munich}
  \country{Germany}
}
\author{Sandra Stankovic}
\email{sandra.stankovic@tum.de}
\orcid{0009-0009-9440-9360}
\affiliation{%
  \institution{Technical University of Munich}
  \country{Germany}
}
\author{Felix Gust}
\email{felix.gust@tum.de}
\affiliation{%
  \institution{Technical University of Munich}
  \country{Germany}
}
\author{Benjamin Lienhard}
\email{benjamin.lienhard@tum.de}
\orcid{0000-0001-5241-4131}
\affiliation{%
  \institution{Technical University of Munich, Department of Physics}
  \country{Germany}
}
\additionalaffiliation{%
  \institution{Walther-Meißner-Institut, Bayerische Akademie der Wissenschaften}
  \city{Garching}
  \country{Germany}
}
\additionalaffiliation{%
  \institution{Munich Center for Quantum Science and Technology}
  \country{Munich, Germany}
}
\author{Pramod Bhatotia}
\email{pramod.bhatotia@tum.de}
\orcid{0000-0002-3220-5735}
\affiliation{%
  \institution{Technical University of Munich}
  \country{Germany}
}

\renewcommand{\shortauthors}{Giortamis et al.}

\begin{abstract}
Quantum error correction (QEC) is the most promising route toward fault-tolerant quantum
computing and, thus, useful quantum computers. QEC operates as a continuous
measure-decode-correct cycle: ancilla qubits are read out, a decoder infers errors from
the resulting syndromes, and corrections are applied before the next round begins. Within
this loop, readout occupies a uniquely critical role, as it is the sole source of ground
truth available to the decoder. Yet readout is also the slowest and most error-prone
operation in the stack, with characteristics that vary across qubits and drift over time;
This complexity propagates directly to the classical control hardware, and in particular
to the FPGA-hosted machine-learning (ML) discriminator that must classify each analog signal into a binary
syndrome outcome. Despite this central role, QEC performance has not yet been studied in depth from
the perspective of readout characteristics, readout length, and their co-design with an
ML discriminator.

We introduce \projectname, an end-to-end benchmarking framework that evaluates qubit-state readout and its impact on QEC performance across real experimentally extracted qubit-state-readout datasets, state-of-the-art ML discriminators, multiple QEC codes, and hardware regimes spanning current to projected devices. Our study reveals three asymmetric findings: The measurement duration can be significantly reduced with nearly no penalty to the logical error rate; The discriminator complexity barely affects the QEC performance, as residual errors are written into device physics rather than the model; and the impact of qubit-state readout on the logical error rate is conditional on where the hardware sits in the QEC landscape, a window that widens as devices mature.
\end{abstract}

\maketitle

\section{Introduction}
Existing quantum computers are inherently noisy: errors accumulate faster than useful computation can proceed, and unfortunately, hardware improvements alone cannot close this gap. As such, quantum error correction (QEC) is currently the \textbf{best route} to \emph{fault-tolerant quantum computing} (FTQC), the mechanism by which quantum processors can reliably execute the large-scale algorithms that promise a transformative advantage in quantum chemistry, materials simulation, drug discovery, and cryptanalysis~\cite{Preskill2018Quantum,Nielsen2010Quantum,Shor1994Algorithms,Motta2023Quantum,Gidney2021How,Bravyi2022Future}. In short, scaling QEC to sufficient code distances is what defines the boundary between today's noisy devices and the fault-tolerant era~\cite{Shor1995Scheme,Terhal2015Quantum,Fowler2012Surface}.

\begin{figure}[t]
    \centering
    \includegraphics[width=\linewidth]{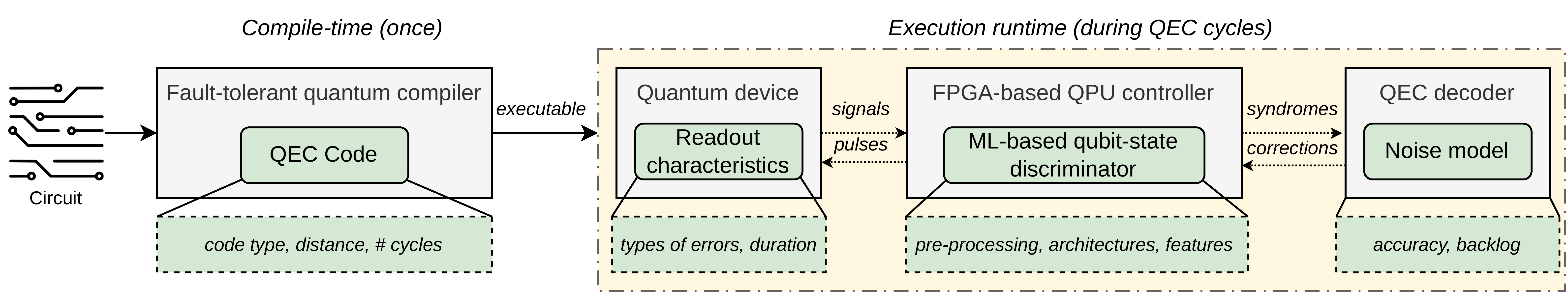}
    \caption{Lifetime of a QEC-protected application. Our framework-based analysis covers the elements in green.}
    \label{fig:qec_pipeline}
\end{figure}

QEC suppresses errors through redundancy: each \emph{logical} qubit is encoded across many \emph{physical} qubits in a way that makes noise detectable and correctable without disturbing the encoded information~\cite{Shor1995Scheme,Fowler2012Surface}. The computation unfolds in discrete \emph{QEC cycles}, as shown in Fig.~\ref{fig:qec_pipeline}: in each cycle, the QPU \emph{control hardware} reads out a set of dedicated ancilla qubits, and produces specific measurement outcomes called \emph{syndromes}; a decoder analyzes the syndrome pattern in real time to infer where and what errors have occurred; and corrections are applied to the data qubits before the next cycle begins~\cite{Terhal2015Quantum,Fowler2012Surface}. This loop runs continuously throughout the computation, keeping the number of errors below an acceptable threshold as long as the physical error rate stays within the code's tolerance.

Within each QEC cycle, qubit-state readout occupies a uniquely central role: it is the \emph{sole source of ground truth} available to the decoder. Every correction decision the decoder makes is based entirely on the binary outcomes of ancilla measurements; there is no other channel through which the classical compute stack can observe errors. This makes readout the critical interface between the quantum processor and the classical control hardware, and its quality propagates directly from QPU physics through the electronics all the way to logical and application-level performance~\cite{Harper2026Characterising}. This propagation runs through two essential stages: the analog readout signal itself and the discriminator that converts it into a binary syndrome bit; we examine each in turn.

First, qubit-state readout is neither simple nor uniform. At the QPU level, each qubit has individual readout characteristics (e.g., coupling strengths or resonator frequencies). This has two implications: (1) readout fidelity is heterogeneous across qubits and can drift across time~\cite{Tannu2019Mitigating,Heinsoo2018Rapid,Thorbeck2024Readout}, (2), fidelity is tunable with respect to measurement duration: a longer readout window accumulates more signal and improves fidelity, while a shorter readout is more challenging and fundamentally bounded by the qubit and resonator properties~\cite{Krantz2019Quantum,Walter2017Rapid,Mallet2009Single}. Critically, current QPUs are calibrated and optimized for \emph{final circuit} measurements rather than the high-rate syndrome readouts that QEC requires during execution~\cite{Harper2026Characterising}. Nonetheless, readout is typically one of the \textbf{slowest and most error-prone operations} in the stack: a single readout takes $1.5$--$2.5\,\mu$s, a $\sim 20$--$40\times$ overhead over the next most expensive operation, and its error rate ($\sim$1--4\%) exceeds gate error by up to an order of magnitude (Table~\ref{tab:readout-errors}, App.~\ref{app:error-rates})~\cite{Yang2022Fpga,Chen2023Transmon,Maurya2023Scaling}.

\begin{wrapfigure}{r}{0.4\textwidth}
    \centering
    \includegraphics[width=\linewidth]{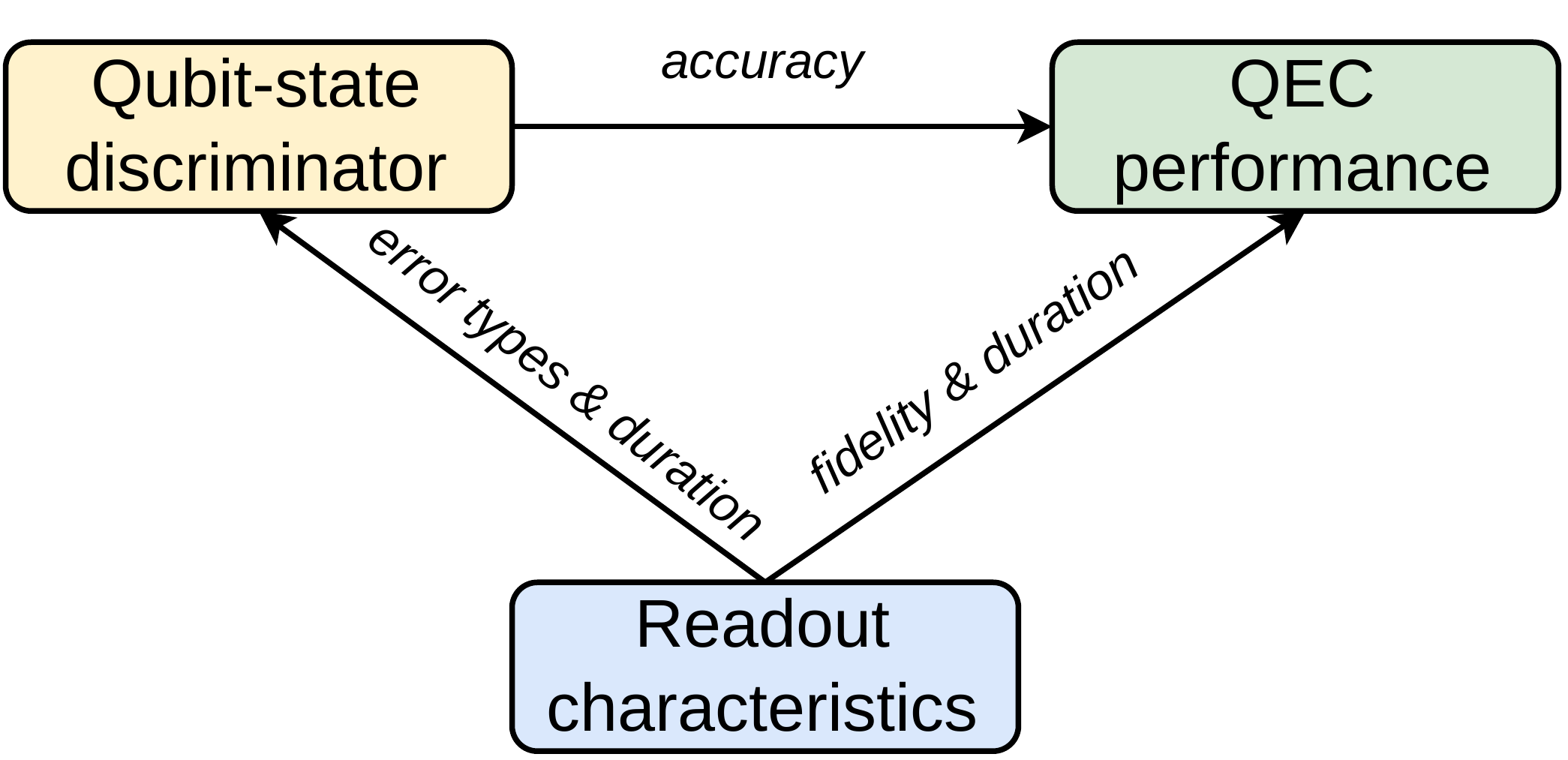}
    \caption{Readout characteristics drive both the qubit-state discriminator and QEC performance, and the discriminator further shapes QEC performance. }
    \label{fig:axes}
\end{wrapfigure}

Second, the complexity of the readout directly propagates to the QPU control hardware (Fig.~\ref{fig:background}): classical controllers, implemented on FPGAs, are tightly constrained in both resources and timing, as they handle the full signal-processing pipeline for the QPU in real time~\cite{Xu2021Qubic,Fruitwala2024Distributed,Tian2025Artery}. Central to this pipeline is a \emph{qubit-state discriminator}, an ML model that classifies the complex analog signal from the QPU into a binary outcome ($|0\rangle$ or $|1\rangle$). The choice of discriminator depends heavily on the readout characteristics of the device, spanning a wide range from simple linear thresholding to neural networks and full transformer architectures~\cite{Magesan2015Machine,Lienhard2022Deep,Maurya2023Scaling,Vora2024Ml,Guo2025Klinq,Guglielmo2025End}. Critically, the accuracy of this classification is not a hardware-internal concern: every misclassified outcome is a corrupted syndrome, directly degrading the information available to the decoder and deteriorating QEC performance, with measurement noise consistently found to play a dominant role in logical gate performance~\cite{Harper2026Characterising}.

To sum up, QEC performance is therefore \textbf{not determined by the quantum hardware alone}, as shown in Fig.~\ref{fig:axes}. At the \emph{hardware level}, per-qubit readout characteristics dictate the design of the FPGA controller and the discriminator it must run. At the \emph{systems level}, measurement duration and discriminator accuracy (and their co-design with the controller) set the quality of the syndromes the decoder receives and, consequently, the QEC- and application-level performance.

\takeawayorange{
\noindent\textbf{Research question.} \textit{How is QEC performance impacted by readout characteristics, measurement duration, and their co-design with the qubit-state discriminator?}}

Answering this question is hard, and not for lack of motivation. Three challenges stand in the way, which we develop in \S\ref{sec:motivation}: \textbf{(1)} readout physics, ML/FPGA discriminator design, and QEC simulation are mature in isolation but run on disconnected toolchains, so closing the loop demands new infrastructure; \textbf{(2)} every discriminator in the literature is evaluated under its own protocol, leaving no standardized basis for comparison; and \textbf{(3)} the resulting design space, spanning measurement durations, discriminator architecture, code family, and hardware regime, is large and coupled rather than separable into independent single-axis studies.

To answer our research question, we introduce \projectname\ (Fig.~\ref{fig:oraqle_overview}), a new end-to-end benchmarking framework for the systematic evaluation of qubit readout and its impact on QEC performance. \projectname\ is general, supporting a wide array of readout datasets and analyses passes, ML discriminator architectures (drawn from a systematic study of state-of-the-art models to ensure a fair and comprehensive evaluation), and QEC codes. \projectname{} is flexible, enabling evaluation accross QPU noise profiles ranging from current to near-term and projected future hardware, and it is modular, allowing researchers to integrate new datasets, discriminator models, and codes as they emerge.

Leveraging \projectname, we structure our evaluation around eight research questions that span three axes (Table~\ref{tab:research_questions}). On the \emph{readout axis}, we use real, experimentally extracted readout traces to characterize how measurement duration and per-qubit characteristics shape the raw signal; these properties propagate downstream and directly influence both discriminator accuracy and QEC performance (\S\ref{sec:readout}). On the \emph{discriminator axis}, we systematically evaluate state-of-the-art ML qubit-state discriminators under identical conditions: the same dataset, the same protocol, reporting accuracy, and FPGA cost across all model families (\S\ref{sec:discriminators}). On the \emph{QEC axis}, we study how readout characteristics and discriminator choice jointly affect the \emph{logical error rate} (the key performance indicator for fault-tolerant quantum computing) across six QEC codes and hardware regimes spanning current to projected devices (\S\ref{sec:qec_impact}). Table~\ref{tab:scope} summarizes the scope of this evaluation: the breadth of dimensions we cover, and the specific, established baselines studied within each.

\begin{table}[t]
    \centering
    \scriptsize
    
    \caption{\projectname's scope: breadth across dimensions, and the specific instances studied within each.}
    \label{tab:scope}
    \renewcommand{\arraystretch}{1.3}
    \resizebox{\columnwidth}{!}{%
    \begin{tabular}{|l|p{0.28\linewidth}|p{0.45\linewidth}|}
        \hline
        \textbf{Dimension} & \textbf{Breadth} & \textbf{Instances studied} \\
        \hline
        Readout data & Truncated- \& sliding-window sweep, information-gain-rate analysis, and per-qubit error-type characterization & The experimentally-extracted IQ trace dataset of Lienhard et al.~\cite{Lienhard2022Reinforcement} (no other raw-trace dataset is publicly available) \\
        \hline
        Discriminators & Six state-of-the-art architectures  & Baseline FNN~\cite{Lienhard2022Deep}, HERQULES~\cite{Maurya2023Scaling}, QubiCML~\cite{Vora2024Ml}, KLiNQ~\cite{Guo2025Klinq}, MCMit Transformer and CNN~\cite{Giortamis2025Mcmit} \\
        \hline
        QEC codes & Six codes spanning all four structural families; different code distances & Surface, Color (topological); Steane (concatenated); BB Gross (qLDPC); Bacon-Shor, Heavy-hex (subsystem) \\
        \hline
        Noise models & Current to projected noise & Current-generation (IBM Heron~r3~\cite{UnknownXXXXIbma}); futuristic, projected \\
        \hline
        QEC simulators & Cross-validated across two tools & ECCentric~\cite{Wierkowska2025Eccentric}, lattice-sim~\cite{Maurya2025Synchronization} \\
        \hline
    \end{tabular}%
    }
\end{table}

\myparagraph{Key findings} QEC performance is shaped by readout in surprising and asymmetric ways: 
\takeawayblue{
\begin{itemize}
\item \textbf{Measurement duration can be cut significantly at virtually zero QEC cost.} Most of the readout signal arrives far earlier than the conventional window assumes; shortening the window yields minimal discrimination accuracy loss, which translates to virtually
no penalty in logical error rate (\S\ref{sec:readout}).
\item \textbf{QEC logical error rate is largely insensitive to discriminator complexity.} The residual errors after classification are written into the device physics, not the model; the largest, most complex discriminator buys almost nothing, the simplest capable
one does not, and the logical error rate reflects this  (\S\ref{sec:discriminators}).
\item \textbf{The impact of readout on QEC is conditional.} Whether readout characteristics translate to a meaningful shift in logical error rate depends on where the device sits in the QEC landscape; that window is only widening as hardware matures (\S\ref{sec:qec_impact}).
\end{itemize}}

\myparagraph{Implications} Our findings provide actionable directives across the quantum computing stack:

\takeawaygreen{
\begin{itemize}
\item \textbf{Use shorter readout windows.} Control-hardware engineers should default readout windows to the point where fidelity plateaus. This reclaims valuable latency at a negligible cost to the assignment error rate (\S\ref{sec:readout}).
\item \textbf{Select the most efficient discriminator.} Discriminator designers should avoid sacrificing FPGA logic and latency for marginal accuracy gains that the underlying device physics cannot support. Readout fidelity should be driven by superior readout hardware rather than oversized models (\S\ref{sec:discriminators}).
\item \textbf{Co-design measurement duration with QEC and noise regimes.} QEC architects should treat measurement duration as a tuning knob only in near-threshold, measurement-limited regimes; otherwise, as an easy latency win rather than a performance lever (\S\ref{sec:qec_impact}).
\end{itemize}}

\myparagraph{Contributions} In summary, we make the following contributions:
\takeawaypurple{
\begin{itemize}
\item \textbf{\projectname: an end-to-end benchmarking framework for qubit readout and its impact on QEC.} \projectname\ supports diverse readout datasets, a systematically selected set of ML qubit-state discriminators, and multiple QEC codes and noise profiles spanning current to projected hardware, enabling rigorous, systematic evaluation across the full readout-to-logical-error-rate pipeline.
\item \textbf{A first systematic and standardized comparison of qubit-state discriminators.} Using experimentally-extracted readout traces and identical experimental conditions, we evaluate state-of-the-art ML discriminators, reporting accuracy, FPGA cost, and the conditions under which model complexity is and is not justified.
\item \textbf{A QEC-level study of readout impact across codes and hardware regimes.} Using \projectname, we quantify how measurement duration and discriminator choice jointly affect the logical error rate across six QEC codes and from current to projected hardware, isolating the regimes where readout is an effective lever on QEC performance.
\item \textbf{Actionable, measurement-grounded guidelines for co-designing readout, discriminator, and QEC.} We distill our findings into concrete design lessons on which factors to prioritize, which to deprioritize, and where the remaining physical and systems limits lie.
\end{itemize}}

\section{Background and Motivation}
\label{sec:background}


\subsection{Quantum Error Correction (QEC)}
\label{subsec:qec}

QEC is the most promising path toward fault-tolerant quantum computing~\cite{Shor1995Scheme,Terhal2015Quantum,Preskill2018Quantum}. Since quantum states decohere rapidly~\cite{Nielsen2010Quantum} and they also cannot be copied~\cite{Wootters1982Single}, most QEC protect a \emph{logical} qubit by distributing its information across many physical qubits and detecting errors through \emph{stabilizer} measurements that reveal an error syndrome without collapsing the encoded state~\cite{Terhal2015Quantum,Knill1997Theory}. Crucially, a stabilizer measurement \emph{is} a qubit readout: the decoder never observes the data qubits directly, only the binary outcomes of ancilla measurements, so a misread ancilla is, to the decoder, \textbf{indistinguishable from a genuine physical error} on the code.
QEC codes differ widely in structure, spanning topological, concatenated, qLDPC, and subsystem families that trade off qubit overhead, connectivity, and threshold differently~\cite{Dennis2002Topological,Knill1996Concatenated,Breuckmann2021Quantum,Bacon2006Operator}; we detail the six representative codes we study in \S\ref{overview:taxonomies}.



\myparagraph{Threshold and overhead} QEC is governed by a \emph{threshold}: once the physical error rate, measurement error included, falls below a code-specific critical value, the \emph{logical error rate (LER)} drops exponentially with code distance~\cite{Fowler2012Surface,Dennis2002Topological}; above it, scaling up the code only makes things worse. Reaching this threshold is costly regardless of code family: a single logical qubit can require thousands of physical qubits~\cite{Fowler2012Surface,Wierkowska2025Eccentric}. The LER is the \emph{yardstick} we use throughout this paper: it is how readout and discriminator choices, made far upstream of the decoder, ultimately register at the code level (\S\ref{sec:qec_impact}).

\begin{figure}[t]
    \centering
    \includegraphics[width=\linewidth]{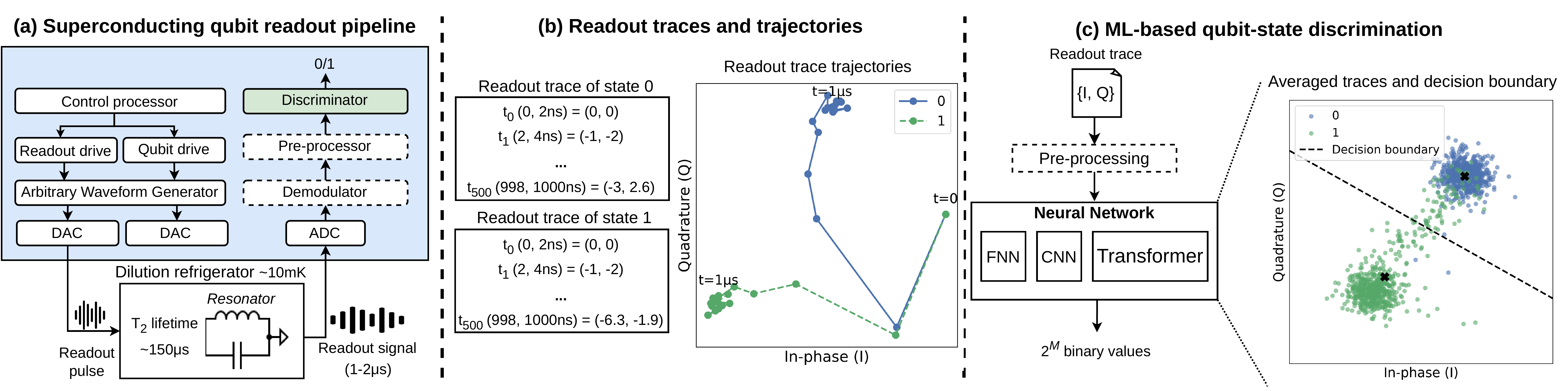}
    \caption{Qubit-state readout pipeline (\S~\ref{sec:background}).\textit{ \textbf{(a)} The signal is processed before a discriminator produces a binary outcome. \textbf{(b)} The trace forms a 2D IQ trajectory over the measurement window. \textbf{(c)} An ML-based discriminator clusters the trajectories and learns a decision boundary to output the qubit state.}}
    \label{fig:background}
\end{figure}

\subsection{Qubit Readout in Quantum Computing Architectures}
\label{subsec:readout}

The stabilizer measurements that the decoder relies on (\S\ref{subsec:qec}) are themselves instances of qubit-state readout, so the syndrome quality QEC depends on is bounded by how well readout itself performs. Qubit-state readout bridges quantum and classical information, extracting a classical outcome from a quantum state after computation~\cite{Yang2022Fpga}. Across qubit platforms, the principle is the same: a measurement returns an analog signal that a discriminator must classify into a binary outcome both \emph{quickly} and \emph{correctly}. We use \emph{readout} in the dispersive regime on superconducting processors (the most mature and widely deployed platform) as a running example~\cite{Acharya2023Multiplexed,Krantz2019Quantum,Mallet2009Single}; the fundamentals carry over to other platforms. In qubit-state readout in the dispersive regime, a resonator coupled to the qubit shifts its frequency depending on the qubit state, so a probe pulse returns a microwave signal whose phase and amplitude encode whether the qubit is in $|0\rangle$ or $|1\rangle$~\cite{Walter2017Rapid,Chen2012Multiplexed}. We detail the underlying physics in App.~\ref{app:readout}.

\myparagraph{Qubit-state readout pipeline} Fig.~\ref{fig:background} illustrates the readout pipeline. The returned signal is amplified, analog demodulated, digitized by an ADC, and then pre-processed (e.g.,\@ via a matched filter) before a discriminator emits a binary outcome (Fig.~\ref{fig:background}a)~\cite{Yang2022Fpga,Macklin2015Near}. The digitally demodulated signal forms a readout \emph{trace}: a sequence of In-Phase (I) and Quadrature (Q) samples tracing a 2D trajectory through the IQ plane over the measurement window (Fig.~\ref{fig:background}b)~\cite{Heinsoo2018Rapid}. The two qubit states drive the trajectory toward two distinct clusters in the IQ plane, and the discriminator assigns the trace to $|0\rangle$ or $|1\rangle$ by learning a decision boundary between them (Fig.~\ref{fig:background}c)~\cite{Magesan2015Machine,Lienhard2022Deep}.

 \myparagraph{Readout dominates the error and latency budget} Typical readout durations of 1.5--2.5\,$\mu s$ are $20$--$40\times$ slower than standard two-qubit gates~\cite{Yang2022Fpga}, and readout error remains up to an order of magnitude higher than single- and two-qubit gate error across all major vendors (Table~\ref{tab:readout-errors}, App.~\ref{app:error-rates}), making it the dominant error source on most processors~\cite{Chen2023Transmon,Acharya2023Multiplexed}. Measurement-induced dephasing and crosstalk in frequency-multiplexed lines further blur the IQ clusters~\cite{Thorbeck2024Readout,Tannu2019Mitigating}, motivating learned discriminators that adapt to this noise without manual threshold tuning (\S\ref{subsec:ml-discrimination}).


\subsection{Machine Learning-Based Qubit-State Discrimination}
\label{subsec:ml-discrimination}

Qubit-state discrimination is the binary classification at the end of the readout pipeline: map a readout trace (the 2D IQ trajectory of Fig.~\ref{fig:background}b) to the state $|0\rangle$ or $|1\rangle$ (Fig.~\ref{fig:background}c)~\cite{Magesan2015Machine,Maurya2023Scaling}. Conventional discriminators are simple and fixed: a linear threshold on the time-integrated IQ point, or a matched filter that projects the trace onto an optimal template before thresholding~\cite{Walter2017Rapid,Maurya2023Scaling}. These are fast, but assume well-separated Gaussian clusters and therefore cannot capture the nonlinear, time-dependent noise of real quantum devices~\cite{Magesan2015Machine,Lienhard2022Deep}.

The state of the art has thus shifted to learned discriminators that fit a decision boundary directly from labeled calibration data, ranging from Gaussian mixture models (GMMs) to feed-forward (FNN), convolutional (CNN), and recurrent (RNN) neural networks~\cite{Magesan2015Machine,Lienhard2022Deep,Maurya2023Scaling,Foumani2024}, improving readout fidelity substantially over simpler baselines~\cite{Maurya2023Scaling}.

\myparagraph{The deployment challenge} Any model must run \emph{on the control hardware} inside the readout loop: inference executes on FPGA electronics under tight latency and area budgets, so discrimination accuracy must be balanced against model size and inference latency to keep real-time feedback viable~\cite{Yang2022Fpga,Duarte2018Fast}. The measurement duration compounds this: many discriminator architectures are duration-dependent, so changing the trace length forces re-training a new model, and some lengths may be unsupportable altogether due to the FPGA's resource budget~\cite{Maurya2023Scaling,Yang2022Fpga}.

\subsection{Motivation: The Qubit-State–Discriminator–QEC Co-Design Space}
\label{sec:motivation}

Qubit-state readout is the decoder's \emph{sole ground truth}: a misclassified syndrome is indistinguishable from a real physical error, and enough of them push the effective error rate past the code threshold, turning QEC into an error amplifier. Measurement duration, discriminator design, and QEC performance are therefore co-dependent, yet they are overwhelmingly studied as independent threads~\cite{Tomita2014Low,Katsuda2024Simulation}. Studying them jointly is not simply a matter of connecting the three communities' tools. Each layer speaks a different language: raw IQ traces, classification accuracy, abstract syndrome error rates, and \textit{no shared infrastructure, benchmark, or design-space methodology bridges them}. This gives rise to three concrete challenges.

%
%

\myparagraph{Challenge 1: no infrastructure connects the three layers} Readout physics, ML/FPGA discriminator design, and QEC simulation are each mature on their own, but their tools were never built to interoperate: a circuit-level QEC simulator ingests an abstract, per-shot error probability, not a raw IQ trace, and a discriminator pipeline stops at classification accuracy, never feeding a decoder. Closing this loop means building the missing glue end-to-end, from raw traces to per-shot discriminator outputs to the correlated measurement errors a QEC simulator can consume, infrastructure that no single prior tool provides.

\myparagraph{Challenge 2: no standardized discriminator comparison} Every state-of-the-art discriminator is evaluated by its own authors, and under its own protocol~\cite{Magesan2015Machine,Lienhard2022Deep,Maurya2023Scaling,Vora2024Ml,Guo2025Klinq}. Before a discriminator choice can be related to anything downstream, every model first has to be reproduced and re-evaluated under one identical protocol, the same traces, the same training recipe, the same hardware cost metrics, an effort substantial enough that no prior work has taken it on.

\myparagraph{Challenge 3: a combinatorial, coupled design space} Measurement duration, discriminator architecture, QEC code family, and hardware noise regime do not vary independently: changing one shifts the feasible range of the others. Exhaustively sweeping this space and cross-validating results across independent QEC simulators to rule out simulator-specific artifacts is an engineering undertaking beyond what any single-axis study has reason to take on.

%
%

\section{Overview}
\label{sec:overview}

We introduce \projectname{} in \S\ref{overview:rscope}, state our hypothesis, and organize the evaluation into eight research questions across three analysis axes. The section then builds the methodology top-down: \S\ref{overview:taxonomies} fixes the discriminator and QEC-code taxonomies that scope our methodology, and \S\ref{sec:methodology} presents the \projectname\ framework as a three-stage read-discriminate-decode pipeline.

\begin{table*}[t]
    \centering
    \caption{ML qubit-state discriminator architectures and feature comparison (\S~\ref{overview:taxonomies}). \textit{Bottom: architectural features and estimated parameter counts, assuming a 5-qubit system with a 1$\mu$s readout window.
    Top: the six discriminators evaluated in \projectname\ in a unified block-diagram notation, using a fixed color code throughout: yellow, purple, red, blue, gray, and green, for linear projections, normalization, feature extraction, sequence aggregation, input and output layers, respectively ($N$ = trace samples, $M$ = qubits or output classes). 
    }}
    \label{tab:model_comparison}
    \includegraphics[width=0.85\textwidth]{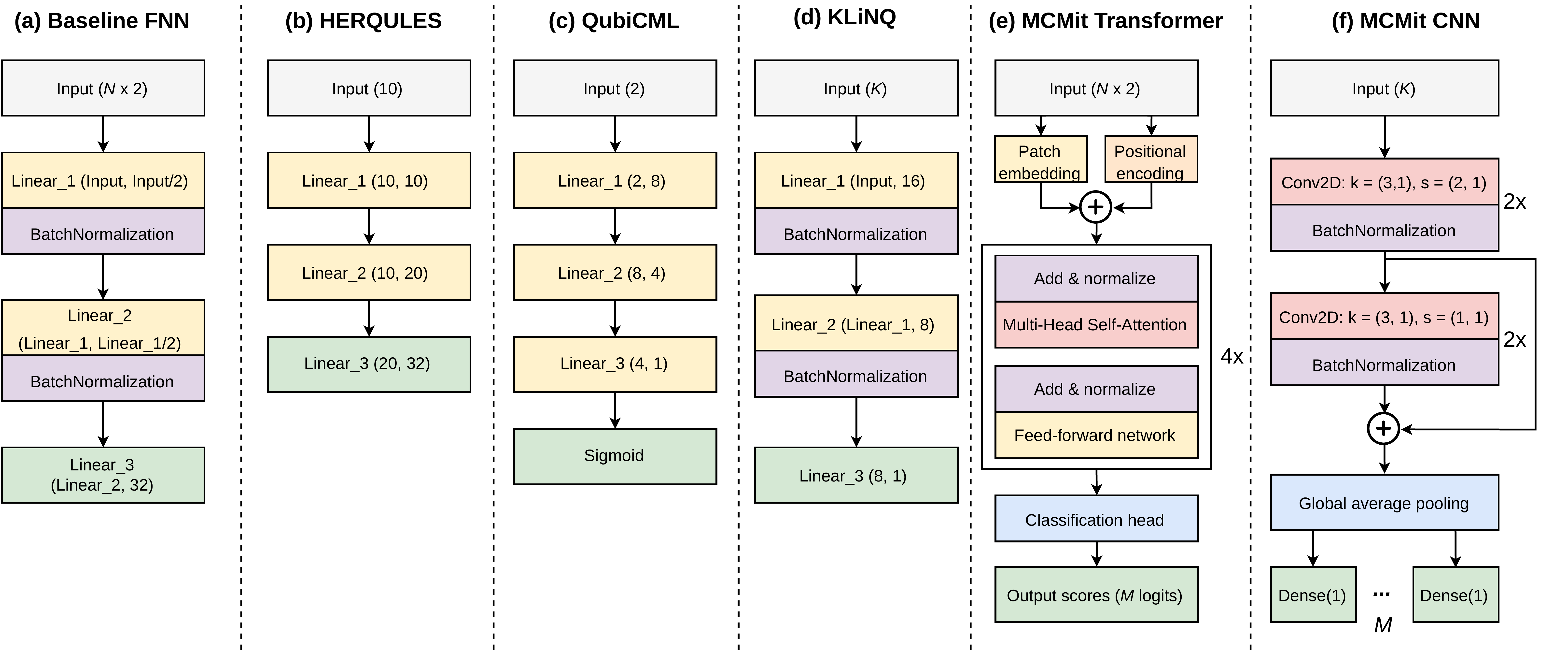}
    \vspace{4pt}
    \resizebox{\textwidth}{!}{%
    \begin{tabular}{l l l l l l l}
        \toprule
        \textbf{Feature} & \textbf{(a) Baseline FNN} & \textbf{(b) HERQULES} & \textbf{(c) QubiCML} & \textbf{(d) KLiNQ} & \textbf{(e) MCMit Transformer} & \textbf{(f) MCMit CNN} \\
        \midrule
        \textbf{Architecture} & FNN  & FNN & FNN & FNN & ViT-style Encoder & Residual CNN \\
        \textbf{Output Type} & Logits (joint 32) & Logits (joint 32) & Probability (Sigmoid) & Logit (per qubit) & Logits (joint 32) & Logits (per qubit) \\
        \textbf{MCM Support} & No & No & Yes & Yes & No & Yes (with padding) \\
        \textbf{Freq-Multiplexed} & Yes & Yes & No\textsuperscript{\dag} & Yes & Yes & Yes \\
        \textbf{Dynamic Length} & No & Yes (via MF) & Yes (via Averaging) & No & Yes (via \texttt{[CLS]}) & Yes (via GAP) \\
        \midrule
        \textbf{Pre-processing} & Flattened trace & MF \& RMF scalars & Demodulated avg. & Trace + Avg + MF & Flattened trace & Downsampling \\
        \textbf{Est. Parameters} & $\sim 635$K & $\sim 1$K & $\sim 65$ (per qubit) & $\sim 9$K\textsuperscript{\ddag} & $\sim 0.1$--$4.8$M\textsuperscript{\S} & $\sim 3$K \\
        \bottomrule
        \multicolumn{7}{l}{\textsuperscript{\dag}\footnotesize{Requires training separate instances for each isolated qubit frequency.}}\\
        \multicolumn{7}{l}{\textsuperscript{\ddag}\footnotesize{KLiNQ operates on a fixed-length (resampled) input, so its parameter count is independent of the readout window.}} \\
        \multicolumn{7}{l}{\textsuperscript{\S}\footnotesize{The MCMit Transformer parameter count varies strongly and non-monotonically with the readout window, from $\sim$0.1M (800\,ns) to $\sim$4.8M (1\,$\mu$s, the reference window).}} \\
    \end{tabular}%
    }
     
\end{table*}

\subsection{Research Scope}
\label{overview:rscope}

\projectname\ is our answer to the three challenges laid out in \S\ref{sec:motivation}: a single framework that bridges measurement, discriminator, and QEC simulation, standardizes discriminator comparison, and makes the resulting design space tractable. We hypothesize that measurement duration and discriminator accuracy are \textit{meaningful, tunable levers} on QEC performance; to test this hypothesis, we structure our evaluation around eight research questions, organized into three axes that mirror the read-discriminate-decode chain (Table~\ref{tab:research_questions}): readout-level analysis, discriminator-level systems tradeoffs, and QEC-level impact. Each axis isolates one link in that chain: how measurement duration and per-qubit properties shape the raw signal; how discriminators convert that signal into a syndrome bit under FPGA latency, logic, and memory constraints; and how the resulting syndrome quality propagates to the logical error rate. Tracing each link separately lets us pinpoint exactly where the hypothesized chain holds and where it breaks down.

\subsection{Qubit-State Discriminator and QEC Code Taxonomies}
\label{overview:taxonomies}

To structure our evaluation, we build on two complementary taxonomies. For ML discriminators, we introduce our own, based on architecture, preprocessing, scalability, and MCM support. For QEC codes, we adopt ECCentric's classification~\cite{Wierkowska2025Eccentric}, which organizes codes into structural families, each with a single representative. 

\myparagraph{ML-based qubit-state discriminators}
We evaluate six state-of-the-art ML-based discriminators. The \emph{Baseline FNN}~\cite{Lienhard2022Deep} is a fully-connected network on raw, frequency-multiplexed IQ traces. \emph{HERQULES}~\cite{Maurya2023Scaling} first compresses each trace into matched-filter (MF) scalars. \emph{QubiCML}~\cite{Vora2024Ml} is a per-qubit FNN on time-averaged IQ signals. \emph{KLiNQ}~\cite{Guo2025Klinq} is a knowledge-distilled compact student FNN. The two sequence-aware \emph{MCMit} models~\cite{Giortamis2025Mcmit} are a Vision Transformer (ViT)-style encoder and a residual convolutional network, both natively handling variable-length traces. Together, they span a wide design space in architecture, preprocessing, parameter count, and MCM support.

\myparagraph{Discriminator architectures}
Table~\ref{tab:model_comparison} (top) depicts all six discriminators in a unified block-diagram notation with a fixed color code, making the routing from raw IQ samples to a decision easy to compare. The four FNN-based models are dominated by projection and normalization blocks; the MCMit Transformer adds attention-based feature extraction with a \texttt{[CLS]}-token aggregation, while the MCMit CNN uses residual convolutional blocks followed by global-average pooling (GAP).

\myparagraph{Discriminator features}
Two design axes in Table~\ref{tab:model_comparison} matter most. \emph{MCM support}: QubiCML, KLiNQ, and the MCMit CNN natively handle mid-circuit measurements, whereas the Baseline FNN, HERQULES, and MCMit Transformer require fixed-length input. \emph{Dynamic-length inference}: all but the Baseline FNN and KLiNQ accept arbitrary trace lengths. Parameter counts span orders of magnitude (from $\sim$65 per qubit for QubiCML to $\sim$4.8M for the MCMit Transformer), directly affecting FPGA resource utilization and inference latency.


\myparagraph{QEC codes}
Following ECCentric's taxonomy~\cite{Wierkowska2025Eccentric}, we evaluate six codes spanning all major structural families. From the \emph{topological} family we take the rotated surface code~\cite{Fowler2012Surface,Dennis2002Topological} and the triangular color code~\cite{Chamberland2020Triangular,Bombin2006Topological}; from the \emph{concatenated} family, the concatenated Steane code~\cite{Steane1996Error,Knill1996Concatenated}; from the \emph{qLDPC} family, the Bivariate Bicycle (BB) Gross code~\cite{Bravyi2024High,Breuckmann2021Quantum}, which reaches a high threshold at far lower qubit overhead than the surface code; and from the \emph{subsystem} family, the Bacon-Shor code~\cite{Bacon2006Operator}, whose single-qubit syndromes are naturally robust to measurement errors, and the heavy-hexagon code~\cite{Chamberland2020Topological}, tailored to IBM's low-degree heavy-hex connectivity. Together, they span the primary trade-offs in qubit overhead, threshold, connectivity, and measurement-noise sensitivity.

\begin{table}[t]
    \centering
    \small
    \caption{Research questions we pursue in this work, and the sections where they are discussed.}
    \label{tab:research_questions}
    \renewcommand{\arraystretch}{1.3}
    \resizebox{\columnwidth}{!}{%
    \begin{tabular}{|c|p{0.9\linewidth}|}
        \hline
        \textbf{Section} & \textbf{Research Question} \\
        \hline
        \multirow{2}{*}{\shortstack{\S~\ref{sec:readout}\\\textit{Readout}}} & \cellcolor{blue!15}\textbf{RQ\#1:} How does measurement duration affect qubit-state assignment fidelity? \\
        \hhline{~|-}
        & \cellcolor{blue!15}\textbf{RQ\#2:} Which readout window is the most critical for achieving high qubit-state assignment fidelity? \\
        \hline
        \multirow{3}{*}{\shortstack{\S~\ref{sec:discriminators}\\\textit{Discriminators}}} & \cellcolor{red!15}\textbf{RQ\#3:} How accurate are state-of-the-art (SOTA) ML-based qubit-state discriminators? \\
        \hhline{~|-}
        & \cellcolor{red!15}\textbf{RQ\#4:} Across the SOTA ML discriminators, what is the tradeoff between model size (parameter count) and discrimination accuracy? \\
        \hhline{~|-}
        & \cellcolor{red!15}\textbf{RQ\#5:} Which readout errors are the hardest to discriminate and why? \\
        \hline
        \multirow{3}{*}{\shortstack{\S~\ref{sec:qec_impact}\\\textit{QEC}}} & \cellcolor{green!15}\textbf{RQ\#6:} How does discriminator accuracy impact logical error rates? \\
        \hhline{~|-}
        & \cellcolor{green!15}\textbf{RQ\#7:} On current hardware, how does measurement duration affect the logical error rate, and how does hardware quality change the payoff? \\
        \hhline{~|-}
        & \cellcolor{green!15}\textbf{RQ\#8:} As hardware improves, does measurement duration become a more important QEC knob? \\
        \hline
    \end{tabular}%
    }
\end{table}

\subsection{The \projectname\ Framework}
\label{sec:methodology}

\begin{figure}
    \centering
    \includegraphics[width=0.85\linewidth]{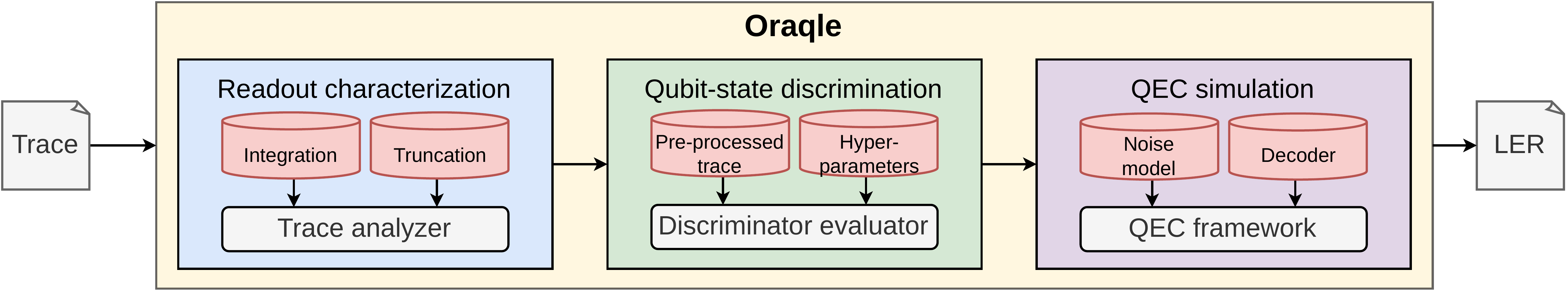}
    \caption{\projectname{} overview. \textit{
    \projectname{} supports diverse trace analysis passes, discriminators, and QEC frameworks. 
    }}
    \label{fig:oraqle_overview}
\end{figure}

\projectname\ (Fig.~\ref{fig:oraqle_overview}) instantiates the QEC read-discriminate-decode cycle as a three-stage evaluation pipeline: real readout traces feed a discriminator stage, whose classification outputs feed a QEC simulation stage that closes the loop with logical error rates. Each stage is self-contained but designed so that its outputs flow directly into the next.

\myparagraph{Stage 1 (Read): Readout characterization}
Since commercial cloud providers \textbf{do not} release raw readout traces publicly, we conduct all readout- and discriminator-level experiments on the dataset of Lienhard et al.~\cite{Lienhard2022Reinforcement}, consistent with all SOTA qubit-state discriminator evaluations. The quantum device is a five-qubit superconducting transmon processor with per-qubit readout resonators\footnote{$T_1$ is in the 12--41\,$\mu$s range}. The dataset contains $\sim$1.6\,M shots, each a time-resolved $(I, Q)$ trajectory sampled at 500\,MSamples/s over a 2\,$\mu$s window and labeled with the prepared basis state, covering all $2^5 = 32$ states with 50,000 traces each. For readout-level analysis (\S\ref{sec:readout}), we additionally employ a simple integration-and-threshold baseline: we trapezoidally integrate the demodulated IQ trace to one $(\bar{I}, \bar{Q})$ point per shot, standardize, and fit a logistic-regression boundary, refit independently at every trace-length truncation so it is always optimally placed for the given SNR.

\myparagraph{Stage 2 (Discriminate): ML discriminator evaluation}
The trace dataset from Stage~1 is the common input to all discriminator evaluations, ensuring a fair, systematic comparison. For every model in \S\ref{sec:discriminators}, we reuse the original authors' training code and recipe, which we obtained for all six models, and additionally run Optuna hyper-parameter optimization to ensure each is reported in a competitive configuration. Per-model training details are deferred to \S\ref{sec:discriminators}.


\myparagraph{Stage 3 (Decode): QEC code simulation}
The per-shot classification outcomes from Stage~2 feed directly into circuit-level noise simulations that close the evaluation loop with logical error rates. We simulate the codes from \S\ref{overview:taxonomies} using two frameworks, ECCentric~\cite{Wierkowska2025Eccentric} and lattice-sim~\cite{Maurya2025Synchronization}, to cross-validate results across different noise models and decoders. For each code, we sweep available distances under two QPU noise profiles: a \textit{current-generation} model calibrated to contemporary hardware, and a \textit{future-generation} model with projected gate and measurement error rates. Per-simulator details and noise parameters are in \S\ref{sec:qec_impact}.

\section{Understanding Qubit Readout}
\label{sec:readout}

This section characterizes the interplay between measurement duration and qubit-state assignment fidelity at the \textit{signal level}, using a model-agnostic integration-and-threshold baseline (ML discriminators are deferred to \S\ref{sec:discriminators}). Pinning the classifier to this baseline attributes any fidelity change to the readout trace itself, yielding a clean relationship that feeds both discriminator design (\S\ref{sec:discriminators}) and QEC performance (\S\ref{sec:qec_impact}). We pursue two objectives: quantifying how trace length affects assignment fidelity (\S\ref{subsec:readout_length}), and identifying the most discriminative time window within the trace (\S\ref{subsec:critical_window}).

\subsection{Measurement Duration Effect on Assignment Fidelity}
\label{subsec:readout_length}

\begin{wrapfigure}{r}{0.5\textwidth}
\centering
\includegraphics[width=0.5\textwidth]{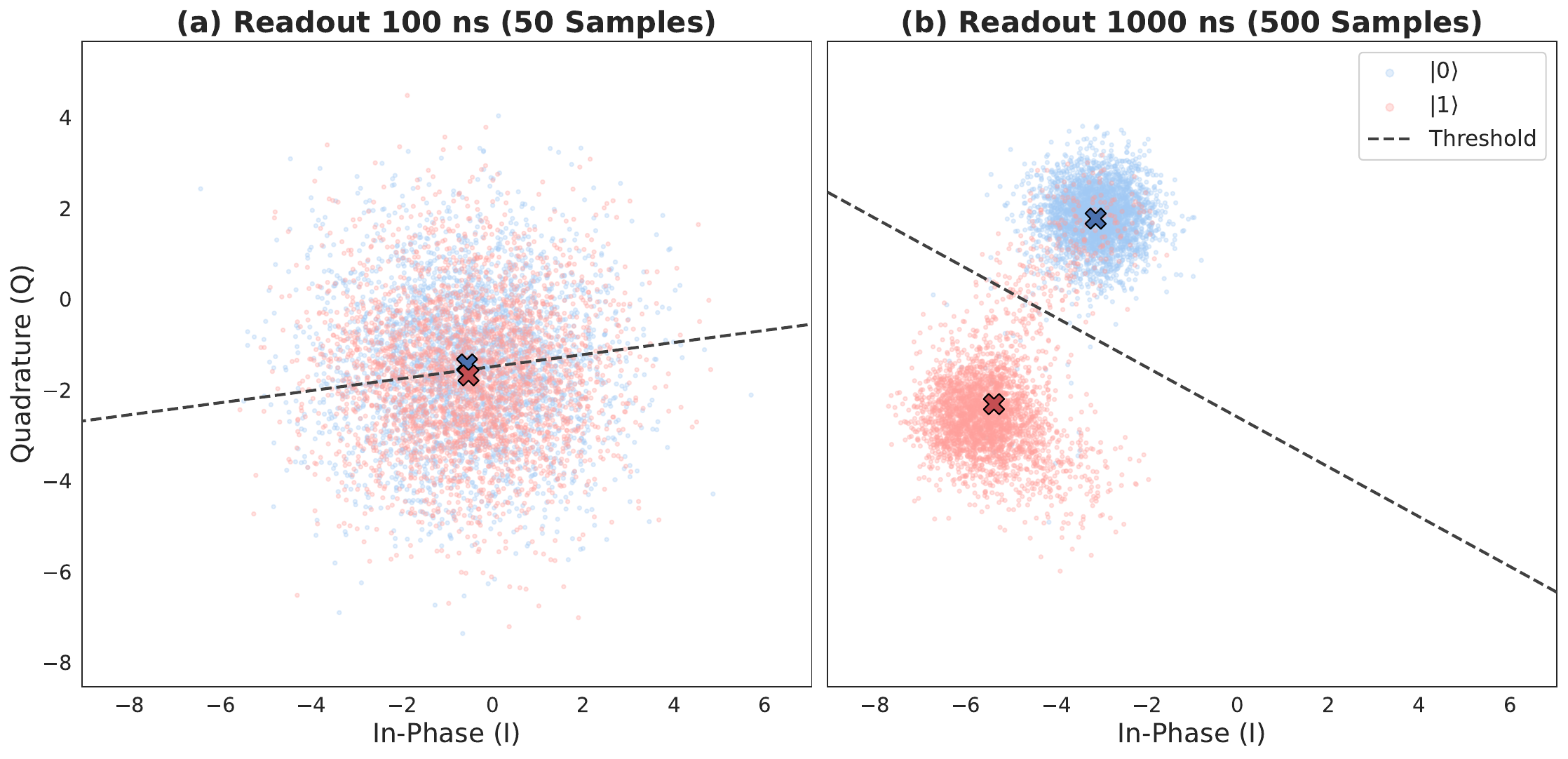}
\caption{Integrated single-shot IQ clouds for Qubit~5: overlapping at a $100$\,ns readout \textbf{(a)} but cleanly threshold-separable at $1000$\,ns \textbf{(b)}.}
\label{fig:iq_clouds}
\end{wrapfigure}

Measurement duration is a design knob: longer integration accumulates more signal energy, thereby achieving higher fidelity, but also incurs more idle decoherence on neighboring data qubits and prolongs QEC cycles. We therefore establish how much of the nominal window can be cut before fidelity drops, and where returns plateau or reverse. Fig.~\ref{fig:iq_clouds} shows what is at stake: for Qubit~5, a $100$\,ns readout collapses the integrated $|0\rangle$ and $|1\rangle$ shots into an overlapping cloud that no linear threshold can separate, whereas the full $1000$\,ns readout pulls the two clusters cleanly apart.

\myparagraph{Research question and hypothesis}
\rqbox{blue}{\textbf{RQ\#1:} \textit{How does measurement duration affect qubit-state assignment fidelity?}}


We hypothesize that fidelity increases with trace length but with diminishing (and eventually \textit{negative}) returns past a saturation point. We expect traces 20--50\% shorter than the nominal $1\,\mu$s window to cost little fidelity, since most discriminative information is concentrated in the resonator's ring-up. Beyond saturation, fidelity should plateau or mildly degrade as the $|0\rangle$/$|1\rangle$ phase separation stops growing while $T_1$ relaxation keeps accumulating.

\myparagraph{Methodology}
Using the baseline and dataset of \S\ref{sec:methodology}, for each truncation point $T \in [0, 1\,\mu\text{s}]$ we trapezoidally integrate the demodulated IQ trace up to $T$, refit the logistic-regression boundary on the resulting $(\bar{I}(T), \bar{Q}(T))$ features, and report per-qubit assignment fidelity and information-gain rate. Refitting independently for every $T$ ensures any fidelity change reflects signal-level effects rather than threshold drift.

\begin{figure} [t]
    \centering
    \includegraphics[width=0.9\linewidth]{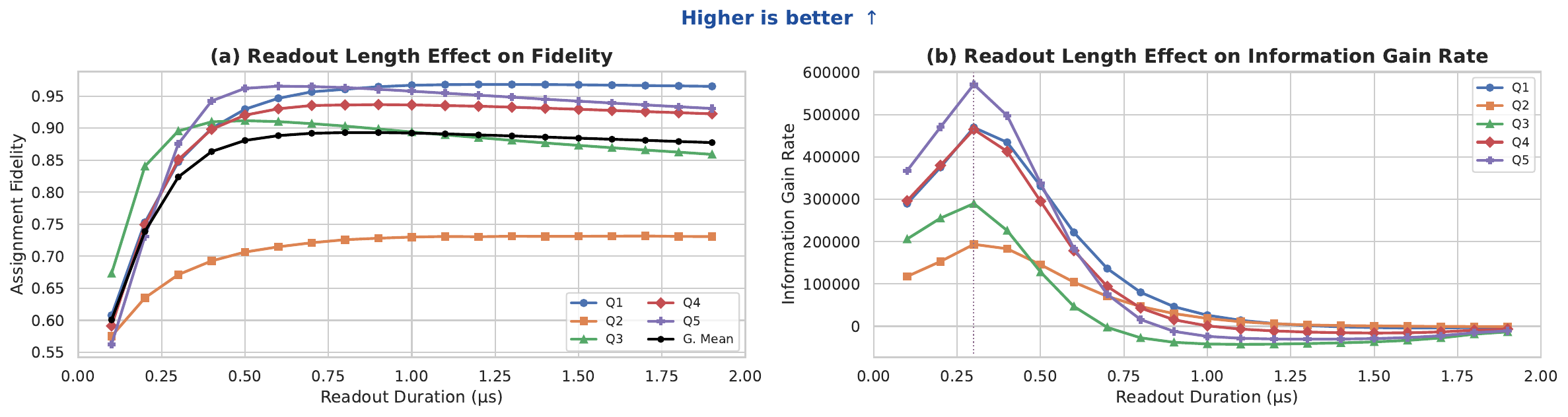}
    \caption{Measurement duration effect on assignment fidelity (\S~\ref{subsec:readout_length}). \textit{(a) Geometric-mean baseline fidelity rises until $t\!\approx\!0.5\,\mu$s, then plateaus (Q1, Q2, Q4) or declines (Q3, Q5). (b) The information-gain rate peaks at $t\!\approx\!0.3\,\mu$s and turns negative beyond $t\!\approx\!1\,\mu$s.}}
    \label{fig:readout_length_impact}
\end{figure}

\myparagraph{Results}
Fig.~\ref{fig:readout_length_impact}(a) sweeps the baseline's per-qubit assignment fidelity over $T \in [0.1, 2.0]\,\mu$s in $0.1\,\mu$s steps. Fidelity rises sharply during the first $0.3$--$0.5\,\mu$s: the geometric mean climbs from $\sim0.60$ at $T=0.1\,\mu$s to $\sim0.86$ at $0.3\,\mu$s and $\sim0.88$ at $0.5\,\mu$s, capturing the bulk of the achievable fidelity within a quarter of the nominal $2\,\mu$s window. Beyond $0.5\,\mu$s the curves split into two regimes: \textit{Q1}, \textit{Q2}, and \textit{Q4} improve slowly to a plateau at $T \approx 1.0\,\mu$s ($\sim0.97$, $\sim0.73$, $\sim0.94$), while \textit{Q3} and \textit{Q5} peak around $0.7$--$1.0\,\mu$s ($\sim0.90$, $\sim0.96$) and then \textit{decrease}, experiencing a relative drop of up to $\sim$40\% by $T=2.0\,\mu$s.

\myparagraph{Analysis}
Fig.~\ref{fig:readout_length_impact}(b) recasts the sweep as the per-qubit \textit{information-gain rate} (the bits per unit time the optimal threshold extracts as $T$ grows) and explains every regime in panel~(a). The rate peaks at $T = 0.3\,\mu$s on all five qubits, declines monotonically, and crosses zero between $T \approx 0.8$ and $1.0\,\mu$s. This maps onto panel~(a): the steep early rise drives the sub-$0.5\,\mu$s gains; the small positive rate from $0.5$--$1.0\,\mu$s gives the slow plateau of Q1, Q2, and Q4; and the negative rate beyond $1.0\,\mu$s makes Q3 and Q5 \textit{lose} fidelity, as each late sample adds more noise than signal to the integrated $(\bar{I}, \bar{Q})$ statistic, which a refit threshold cannot recover.

\takeaway{\textbf{Takeaway \#1: Measurement duration can be reduced.} Assignment fidelity saturates by $T \approx 1\,\mu$s on every qubit and \textit{actively loses} ground past that point. In practice, $\sim$0.5--1.0\,$\mu$s of trace captures nearly all of the discriminative information; every additional nanosecond we spend integrating is latency that the FPGA controller and the QEC cycle no longer get to reclaim.}

\subsection{Identifying the Critical Readout Window}
\label{subsec:critical_window}

Beyond the length–fidelity trade-off: is discriminative information spread evenly across the trace, or concentrated in a specific window?

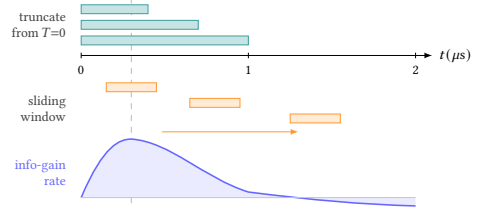
\begin{wrapfigure}{r}{0.5\textwidth}
\centering
\vspace{-\intextsep}
\resizebox{0.45\textwidth}{!}{%
\begin{tikzpicture}[font=\scriptsize,>=latex]
  \draw[gray!55,dashed] (0.9,-0.1) -- (0.9,3.55);
  \node[anchor=east,align=right,black!75] at (-0.15,3.1) {truncate\\from $T{=}0$};
  \foreach \L/\y in {1.2/3.30, 2.1/3.02, 3.0/2.74}{
     \fill[teal!22] (0,\y) rectangle (\L,\y+0.16);
     \draw[teal!75] (0,\y) rectangle (\L,\y+0.16);
  }
  \draw[->] (0,2.55) -- (6.3,2.55) node[right,black]{$t\,(\mu\mathrm{s})$};
  \foreach \x/\t in {0/0, 3/1, 6/2}{
     \draw (\x,2.50)--(\x,2.60);
     \node[below] at (\x,2.50) {\t};
  }
  \node[anchor=east,align=right,black!75] at (-0.15,1.6) {sliding\\window};
  \foreach \c/\y in {0.9/1.90, 2.4/1.62, 4.2/1.34}{
     \fill[orange!16] (\c-0.45,\y) rectangle (\c+0.45,\y+0.16);
     \draw[orange!80] (\c-0.45,\y) rectangle (\c+0.45,\y+0.16);
  }
  \draw[->,orange!80] (1.45,1.18) -- (3.9,1.18);
  \fill[blue!8]
     (0,0) .. controls (0.4,0.95) and (0.7,1.05) .. (0.9,1.05)
           .. controls (1.6,1.0) and (2.2,0.35) .. (3.0,0.1)
           .. controls (4.2,-0.05) and (5.0,-0.12) .. (6,-0.15)
           -- (6,0) -- (0,0) -- cycle;
  \draw[blue!60,thick]
     (0,0) .. controls (0.4,0.95) and (0.7,1.05) .. (0.9,1.05)
           .. controls (1.6,1.0) and (2.2,0.35) .. (3.0,0.1)
           .. controls (4.2,-0.05) and (5.0,-0.12) .. (6,-0.15);
  \draw[gray!45] (0,0)--(6,0);
  \node[anchor=east,align=right,blue!60] at (-0.15,0.45) {info-gain\\rate};
\end{tikzpicture}%
}
\caption{Truncation keeps the high-information trace onset; a fixed-width window slid later walks off the information-gain peak ($\sim$0.3\,$\mu$s, dashed).}
\vspace{-20pt}
\label{fig:window_schematic}
\end{wrapfigure}

\myparagraph{Research question and hypothesis}
\rqbox{blue}{\textbf{RQ\#2:} \textit{Which readout window is the most critical for achieving high qubit-state assignment fidelity?}}

RQ\#1 showed the per-bit information-gain rate peaks around $T \approx 0.3\,\mu$s and drops rapidly thereafter, suggesting the discriminative content is concentrated in time rather than spread uniformly (Fig.~\ref{fig:window_schematic}). We therefore hypothesize a critical $\sim$0.2--0.5\,$\mu$s window in the early trace carries the bulk of the information, so a discriminator restricted to it should approach full-trace fidelity, i.e., large portions of the captured signal can be safely dropped without a meaningful penalty.

\begin{figure} [t]
    \centering
    \includegraphics[width=0.9\linewidth]{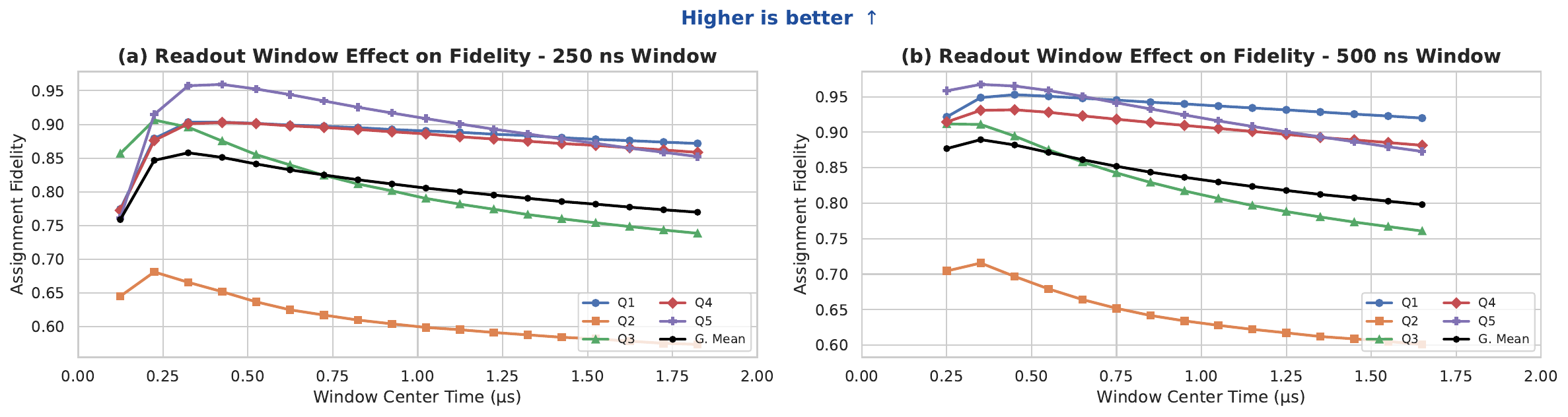}
    \caption{Read window effect on assignment fidelity (\S~\ref{subsec:critical_window}). \textit{
    \textbf{(a)} A $250\,$ns sliding window peaks at $\sim$0.86 fidelity near a $0.3$--$0.4\,\mu$s center and decays monotonically thereafter.
    \textbf{(b)} The $500\,$ns window has the same shape with a higher ceiling ($\sim$0.89 peak). In both, the optimum is anchored near the pulse onset.
    }
    }
    \label{fig:readout_window_impact}
\end{figure}

\myparagraph{Methodology}
We reuse the integration-and-threshold baseline of \S\ref{sec:methodology}, now applied to fixed-width \textit{sliding} windows instead of onset-anchored truncations (Fig.~\ref{fig:window_schematic}). For widths of $250\,$ns and $500\,$ns, we slide each window across the full $2\,\mu$s trace, integrate the demodulated IQ samples inside it, refit the logistic-regression boundary at every position, and report per-qubit and geometric-mean fidelity versus window center.

\myparagraph{Results}
For both window widths, fidelity peaks when the window is centered around $0.3$--$0.4\,\mu$s and decays monotonically as it slides later (Fig.~\ref{fig:readout_window_impact}). The $250\,$ns window swings from $\sim$0.86 at the peak to $\sim$0.77 at a $1.8\,\mu$s center, a $\sim$39\% relative drop for the same integration time, depending only on \textit{where} it sits; the $500\,$ns window has the same shape with a higher ceiling ($\sim$0.89). The per-qubit ordering matches Fig.~\ref{fig:readout_length_impact} (Q1 and Q5 strongest, Q2 weakest) at every position.

\myparagraph{Analysis}
These results refine, rather than contradict, RQ\#1. Although per-bit information gain peaks at $T = 0.3\,\mu$s, we \textit{cannot} exploit this by isolating a narrow window around it: the $250\,$ns window peaks $\sim$21\% (relative) below the $\sim$0.89 plateau of a same-onset truncation, and even the $500\,$ns window never exceeds it. The early samples, though not individually informative, establish the SNR baseline against which later samples are discriminated, so dropping the first $\sim$200--300\,ns hurts. The implication for \S\ref{sec:discriminators}: the discriminator input must stay anchored at $T = 0$, obtained by \textit{truncating} from the start, not by sliding a window.

\takeaway{\textbf{Takeaway \#2: The critical readout window starts at $\mathbf{T = 0}$.} The first $\sim$0.5\,$\mu$s of the trace carries the bulk of the discriminative information, yet we cannot \textit{slide} a narrow window to reduce a discriminator's input: any delay diminishes the fidelity. A shorter discriminator input must therefore be obtained by \textit{truncating} the trace from $T = 0$, not by skipping the early samples.}
\section{Characterization of ML-Based Qubit-State Discriminators}
\label{sec:discriminators}

Having related readout length to accuracy with the simple baseline (\S\ref{sec:readout}), we now turn to the state-of-the-art ML discriminators, whose design space is large and diverse with no consensus on the best architecture for FPGA-deployed readout (\S\ref{overview:taxonomies}). We compare them on the same dataset along three axes: \textit{(i)} accuracy on the full $1\,\mu$s trace and on much shorter traces (\S\ref{subsec:discriminator_perforamnce}); \textit{(ii)} the trade-off between model complexity (parameters, latency, FPGA resources) and accuracy (\S\ref{subsec:discriminator-tradeoffs}); and \textit{(iii)} the conditions under the best models fail (\S\ref{subsec:failures}).

\subsection{Discriminator Performance}
\label{subsec:discriminator_perforamnce}

\begin{figure}[t]
\centering
\resizebox{0.8\linewidth}{!}{%
\begin{tikzpicture}[font=\scriptsize,>=latex,
   sub/.style={black!72}, dim/.style={draw=gray!45,rounded corners=1pt,inner sep=1.6pt,fill=gray!10,text=black!80}]
\begin{scope}[xshift=0cm]
  \node[font=\bfseries] at (1.27,1.22) {raw trace};
  \draw[gray!45] (-0.05,0)--(2.6,0);
  \foreach \x/\h in {0/0.25,0.22/0.55,0.44/0.40,0.66/0.72,0.88/0.58,1.10/0.82,1.32/0.63,1.54/0.76,1.76/0.52,1.98/0.66,2.20/0.46,2.42/0.58}{
     \draw[blue!55] (\x,0)--(\x,\h); \fill[blue!55] (\x,\h) circle(0.7pt);}
  \node[sub] at (1.27,-0.36) {keep every sample};
  \node[dim] at (1.27,-0.68) {$N$ samples};
\end{scope}
\begin{scope}[xshift=3.3cm]
  \node[font=\bfseries] at (1.27,1.22) {matched filter};
  \draw[gray!45] (-0.05,0)--(2.6,0);
  \draw[orange!80,thin] (0,0.30) .. controls (0.6,0.98) and (1.5,0.86) .. (2.42,0.24);
  \foreach \x/\h in {0/0.10,0.22/0.33,0.44/0.32,0.66/0.68,0.88/0.58,1.10/0.78,1.32/0.54,1.54/0.53,1.76/0.29,1.98/0.26,2.20/0.14,2.42/0.12}{
     \draw[blue!55] (\x,0)--(\x,\h); \fill[blue!55] (\x,\h) circle(0.7pt);}
  \node[sub] at (1.27,-0.36) {weight by optimal kernel};
  \node[dim] at (1.27,-0.68) {$1$ to $N$ values};
\end{scope}
\begin{scope}[xshift=6.6cm]
  \node[font=\bfseries] at (1.27,1.22) {downsampling};
  \draw[gray!45] (-0.05,0)--(2.6,0);
  \foreach \x/\h in {0.1/0.30,0.85/0.72,1.6/0.62,2.35/0.55}{
     \draw[blue!55] (\x,0)--(\x,\h); \fill[blue!55] (\x,\h) circle(1.1pt);}
  \node[sub] at (1.27,-0.36) {keep every $k$-th sample};
  \node[dim] at (1.27,-0.68) {$N/k$ samples};
\end{scope}
\begin{scope}[xshift=9.9cm]
  \node[font=\bfseries] at (1.27,1.22) {integration};
  \draw[gray!45] (-0.05,0)--(2.6,0);
  \foreach \x/\h in {0/0.25,0.5/0.6,1.0/0.7,1.5/0.5,2.0/0.55}{ \draw[blue!18] (\x,0)--(\x,\h);}
  \draw[->,gray!60] (0.7,0.55) -- (1.13,0.5);
  \draw[->,gray!60] (1.8,0.55) -- (1.41,0.5);
  \fill[blue!70] (1.27,0.5) circle(1.7pt);
  \node[sub] at (1.27,-0.36) {average over time};
  \node[dim] at (1.27,-0.68) {$1$ point $(\bar{I},\bar{Q})$};
\end{scope}
\node[black!80] at (0.9,-1.04) {high};
\draw[->,black!55,thick] (1.5,-1.04)--(11.0,-1.04);
\node[black!80] at (11.6,-1.04) {low};
\node[black!85] at (6.25,-1.34) {temporal information retained};
\end{tikzpicture}%
}
\caption{Readout-trace preprocessing modes a discriminator can consume, ordered by the temporal information they retain (\S~\ref{subsec:discriminator_perforamnce}). \textit{From raw samples ($N$ values) through matched filtering and downsampling to full integration (one $(\bar I,\bar Q)$ point); the modes compose, and time-truncation (\S~\ref{sec:readout}) applies on top of any of them.}}
\label{fig:preprocessing}
\end{figure}
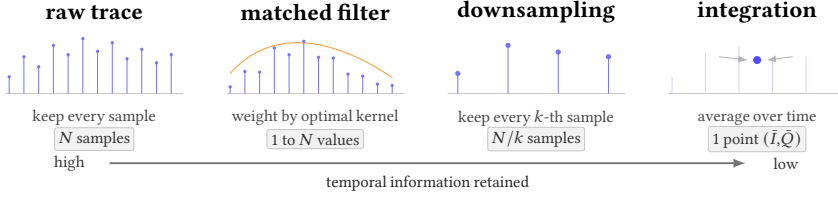

The discriminators vary not only in architecture but in \emph{how much} of the raw trace they expose to the classifier: the demodulated trace can be fed near-verbatim, reweighted by a matched filter, downsampled, or collapsed into a single integrated $(\bar{I},\bar{Q})$ point, each step discarding more temporal structure (Fig.~\ref{fig:preprocessing}). This information content is the primary lever on per-shot fidelity, which in turn directly bounds the logical error rate any QEC code built on these discriminators can achieve (\S\ref{sec:qec_impact}), potentially making even small accuracy differences consequential.

\myparagraph{Research question and hypothesis}
\rqbox{red}{\textbf{RQ\#3:} \textit{How accurate are state-of-the-art ML-based qubit-state discriminators?}}

We hypothesize that ML-based discriminators beat the integration-and-threshold baseline by exploiting temporal dynamics, a static integration discards: e.g., sequence-aware models recognizing the IQ-trajectory bend of mid-trace $T_1$ or excitation events.

\myparagraph{Methodology}
We train all models on the same dataset and evaluate them with the same protocol. For each model we follow its original training recipe; absent a specified split, we use an $80/20$ split of the Lienhard et al.\ dataset (\S\ref{sec:methodology}). Since the duration sweep varies the readout window, we retrain every model from scratch at each duration.

\myparagraph{Results}
Table~\ref{tab:fidelity} (each model's $1000$\,ns row) gives full-$1\,\mu$s fidelity; we measure error reduction as $\frac{(\text{better}-\text{worse})}{(100-\text{worse})}$. \textit{Qubit~2} limits every model ($\sim$0.73--0.75 regardless of architecture), a hardware (not algorithmic) gap, so we report both the five-qubit geometric mean ($F_{5Q}$) and a four-qubit mean ($F_{4Q}$) excluding it. The neural discriminators cluster tightly ($F_{5Q}=0.901$--$0.911$), with the \emph{large} (MCMit-T, Baseline FNN) and \emph{small} (MCMit-CNN, HERQULES, KLiNQ) groups essentially tied; MCMit-CNN even leads on $F_{4Q}$ ($0.957$) at $\sim$3k parameters, so the multi-million-parameter models buy no accuracy. \textit{QubiCML} ($0.893$) is the outlier, no better than the linear baseline ($F_{5Q}=0.892$, Table~\ref{tab:fidelity_linear}), because it integrates each trace to a single $(\bar I,\bar Q)$ point and discards the temporal information before the network sees it. Once a discriminator keeps the raw or matched-filtered trace it reaches essentially the same per-shot fidelity, so the leaders separate on \emph{footprint}, not accuracy (\S\ref{subsec:discriminator-tradeoffs}).

\myparagraph{Analysis}
The duration sweep in Table~\ref{tab:fidelity} adds two points. First, shorter readout is often \emph{better}, not just faster: on the high-SNR qubits the integration- and distillation-based models (QubiCML, KLiNQ) peak below $1\,\mu$s, as extra length accumulates $T_1$ events that corrupt the trajectory faster than added signal helps (\S\ref{subsec:failures}). Second, truncation to $600$\,ns is cheap: every model loses only $\sim$10\% ($F_{5Q}$ $0.901$--$0.911\!\to\!0.894$--$0.901$), reclaiming $40\%$ of the window for a cost the QEC logical error rate barely registers (\S\ref{sec:qec_impact}). Only at the aggressive $200$\,ns point does accuracy break down; models collapse to a $0.738$--$0.767$ band and HERQULES drops furthest (Qubit~2 to coin-flip $0.513$).

\takeaway{\textbf{Takeaway \#3: Truncate to the fidelity plateau.} The readout window can be cut to $\sim$600\,ns for only a $\sim$10\% rise in per-shot error, reclaiming latency that directly shortens the QEC cycle. The gains are confined to the well-separated qubits, while Qubit~2 sits at $\sim$0.69--0.75 for \emph{every} model across the $400$--$1000$\,ns range, a hardware wall no architecture or window length moves (\S\ref{subsec:failures}).}
  
{ \scriptsize
\begin{longtable}{ll rrrrr rr}
\caption{Per-qubit readout fidelity vs.\ trace duration. \textbf{Bold} = best duration for that qubit within a model; \colorbox{blue!12}{shaded} = highest fidelity for that qubit across all configurations. $F_{5Q}$/$F_{4Q}$ are the geometric means over all five qubits / the four excluding the limiting Qubit~2.}
\label{tab:fidelity}\\
\toprule
Model & Duration & Qubit 1 & Qubit 2 & Qubit 3 & Qubit 4 & Qubit 5 & $F_{5Q}$ & $F_{4Q}$ \\
\midrule
\endfirsthead
\toprule
Model & Duration & Qubit 1 & Qubit 2 & Qubit 3 & Qubit 4 & Qubit 5 & $F_{5Q}$ & $F_{4Q}$ \\
\midrule
\endhead
\multirow{5}{*}{Baseline FNN} & 1000\,ns & \textbf{0.968} & \textbf{0.749} & \textbf{0.937} & \textbf{0.945} & 0.969 & \textbf{0.909} & \textbf{0.955} \\
 & 800\,ns & 0.965 & 0.741 & 0.936 & 0.943 & 0.969 & 0.906 & 0.953 \\
 & 600\,ns & 0.953 & 0.730 & 0.933 & 0.936 & \textbf{0.969} & 0.899 & 0.948 \\
 & 400\,ns & 0.915 & 0.705 & 0.923 & 0.910 & 0.958 & 0.877 & 0.926 \\
 & 200\,ns & 0.784 & 0.648 & 0.866 & 0.777 & 0.774 & 0.767 & 0.799 \\
\midrule
\multirow{5}{*}{HERQULES\footnotemark} & 1000\,ns & \textbf{0.969} & \textbf{0.740} & \textbf{0.924} & \textbf{0.944} & \textbf{0.970} & \textbf{0.905} & \textbf{0.951} \\
 & 800\,ns & 0.964 & 0.733 & 0.923 & 0.940 & 0.969 & 0.901 & 0.949 \\
 & 600\,ns & 0.950 & 0.720 & 0.922 & 0.932 & 0.969 & 0.894 & 0.943 \\
 & 400\,ns & 0.906 & 0.695 & 0.917 & 0.899 & 0.945 & 0.867 & 0.917 \\
 & 200\,ns & 0.751 & 0.513 & 0.835 & 0.737 & 0.720 & 0.702 & 0.759 \\
\midrule
\multirow{5}{*}{QubiCML} & 1000\,ns & \textbf{0.967} & \textbf{0.732} & 0.894 & 0.938 & 0.956 & 0.893 & 0.938 \\
 & 800\,ns & 0.963 & 0.725 & 0.903 & \textbf{0.938} & 0.963 & \textbf{0.893} & \textbf{0.941} \\
 & 600\,ns & 0.947 & 0.715 & 0.912 & 0.930 & \textbf{0.966} & 0.889 & 0.939 \\
 & 400\,ns & 0.899 & 0.693 & \textbf{0.915} & 0.896 & 0.944 & 0.864 & 0.913 \\
 & 200\,ns & 0.753 & 0.635 & 0.840 & 0.747 & 0.729 & 0.738 & 0.766 \\
\midrule
\multirow{5}{*}{KLiNQ} & 1000\,ns & \textbf{0.968} & \textbf{0.744} & 0.926 & 0.933 & 0.956 & \textbf{0.901} & 0.946 \\
 & 800\,ns & 0.964 & 0.737 & 0.926 & \textbf{0.933} & 0.961 & 0.900 & \textbf{0.946} \\
 & 600\,ns & 0.952 & 0.725 & \textbf{0.928} & 0.925 & \textbf{0.966} & 0.894 & 0.943 \\
 & 400\,ns & 0.914 & 0.703 & 0.926 & 0.901 & 0.957 & 0.875 & 0.924 \\
 & 200\,ns & 0.783 & 0.647 & 0.866 & 0.775 & 0.773 & 0.766 & 0.798 \\
\midrule
\multirow{5}{*}{MCMit-T} & 1000\,ns & \textbf{0.969} & \cellcolor{blue!12}\textbf{0.749} & 0.941 & \textbf{0.946} & 0.970 & \cellcolor{blue!12}\textbf{0.911} & \textbf{0.956} \\
 & 800\,ns & 0.966 & 0.743 & \textbf{0.941} & 0.944 & 0.970 & 0.909 & 0.955 \\
 & 600\,ns & 0.953 & 0.730 & 0.938 & 0.937 & \textbf{0.970} & 0.901 & 0.949 \\
 & 400\,ns & 0.915 & 0.706 & 0.929 & 0.909 & 0.958 & 0.878 & 0.928 \\
 & 200\,ns & 0.784 & 0.648 & 0.866 & 0.777 & 0.775 & 0.767 & 0.800 \\
\midrule
\multirow{5}{*}{MCMit-CNN} & 1000\,ns & \cellcolor{blue!12}\textbf{0.970} & \textbf{0.746} & \cellcolor{blue!12}\textbf{0.942} & \cellcolor{blue!12}\textbf{0.947} & 0.970 & \textbf{0.910} & \cellcolor{blue!12}\textbf{0.957} \\
 & 800\,ns & 0.965 & 0.737 & 0.941 & 0.944 & 0.970 & 0.907 & 0.955 \\
 & 600\,ns & 0.953 & 0.725 & 0.938 & 0.937 & \cellcolor{blue!12}\textbf{0.970} & 0.900 & 0.949 \\
 & 400\,ns & 0.907 & 0.697 & 0.926 & 0.901 & 0.951 & 0.871 & 0.921 \\
 & 200\,ns & 0.782 & 0.646 & 0.864 & 0.774 & 0.770 & 0.764 & 0.796 \\
\bottomrule
\end{longtable} \vspace{-10pt}
}
\footnotetext{\herqulesnote}

\subsection{ML-Based Discriminator Complexity and Tradeoffs}
\label{subsec:discriminator-tradeoffs}
The accuracy ranking above tells only half the story for FPGA-deployed readout. The discriminator competes for a strict budget of on-chip logic, memory, and latency. Accuracy gains (\S\ref{subsec:discriminator_perforamnce}) are therefore useful only if the model is small and fast enough to deploy alongside the control stack, which motivates an accuracy-vs.-cost analysis of the SOTA suite.

\rqbox{red}{\textbf{RQ\#4:} \textit{Across the state-of-the-art ML-based qubit-state discriminators, what is the tradeoff between model size (parameter count) and discrimination accuracy?}}

\myparagraph{Methodology}
For each discriminator, we plot discrimination accuracy ($F_{5Q}$) against the number of trainable parameters to expose the size-versus-accuracy trade-off, at two readout windows: the full $1\,\mu$s trace and the truncated $600$\,ns window. To ground it in hardware cost, we also synthesize the discriminators and report measured FPGA resource utilization and inference latency (Table~\ref{tab:fpga_resources}).

\begin{figure}[t]
\centering
\includegraphics[width=0.8\linewidth]{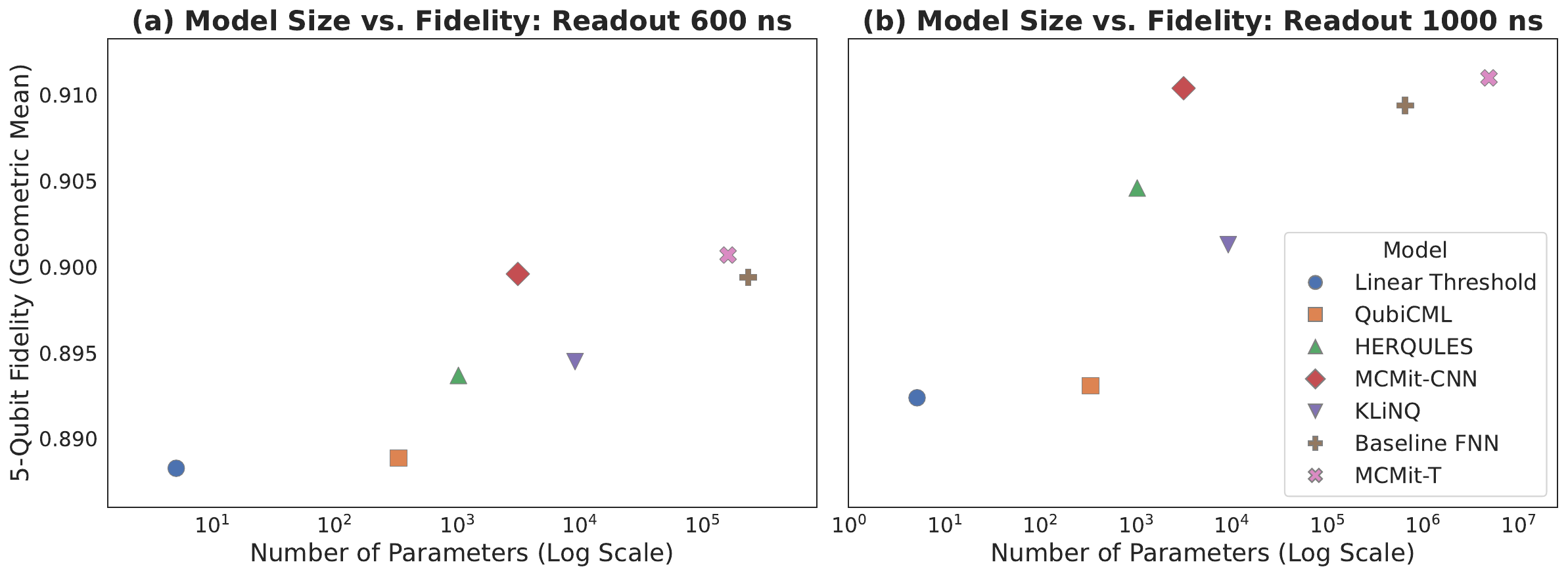}
\caption{Qubit-state discriminator fidelity vs.\ trainable-parameter count (log scale) at \textbf{(a)} $600$\,ns and \textbf{(b)} $1\,\mu$s readout (\S~\ref{subsec:discriminator-tradeoffs}). \textit{MCMit-CNN and HERQULES surpass the orders of magnitude larger models.}}
\label{fig:pareto}
\end{figure}

\myparagraph{Results}
Fig.~\ref{fig:pareto} plots fidelity against parameter count at both windows. The models span five orders of magnitude in size (from the $\sim$5-parameter Linear threshold to the $\sim$5M-parameter MCMit-T), yet only $\sim$18\% in $F_{5Q}$ error reduction. The \emph{small} models define the frontier: MCMit-CNN ($\sim$3k params) matches the multi-million-parameter MCMit-T and Baseline FNN on $F_{5Q}$ and leads the suite on $F_{4Q}$ ($0.957$), with HERQULES ($\sim$1k) just behind; the largest models buy nothing.

\begin{table}[t]
\centering
\footnotesize
\setlength{\tabcolsep}{4pt}
\caption{FPGA cost of the deployable discriminators on the QubiC ZCU216 RFSoC (\texttt{xczu49dr}, $500$\,MHz), for full five-qubit readout. LUT/FF/DSP are \% of available device resources; latency is per inference.}
\label{tab:fpga_resources}
\begin{tabular}{lcccc}
\toprule
\textbf{Discriminator} & \textbf{LUT} & \textbf{FF} & \textbf{DSP} & \textbf{Latency} \\
\midrule
QubiCML\textsuperscript{\dag} & 3\% & 3\% & 5\% & $38$\,ns \\
HERQULES & 6\% & 9\% & 17\% & $38$\,ns \\
MCMit CNN & 29\% & 16\% & 20\% & $74$\,ns \\
KLiNQ\textsuperscript{\ddag} & 70\% & 49\% & 69\% & $42$\,ns \\
\bottomrule
\multicolumn{5}{l}{\textsuperscript{\dag}\footnotesize Per-qubit model, replicated across all five qubits.}\\
\multicolumn{5}{l}{\textsuperscript{\ddag}\footnotesize Per-qubit; three qubits use the small student, two the larger.}\\
\end{tabular}
\end{table}

\myparagraph{Analysis}
Capacity is not the bottleneck: once a model retains the raw trace, a few thousand parameters saturate the achievable fidelity, so the giants' extra $100$--$1000\times$ weights buy nothing. What sets parameter count is the input dimension the first layer must absorb, kept small by two levers: preprocessing that shrinks the input (HERQULES, KLiNQ) or weight sharing that decouples it from length (MCMit-CNN); only a long raw trace into a dense layer (Baseline FNN, MCMit-T) inflates. This footprint story holds on real hardware (Table~\ref{tab:fpga_resources}): the Baseline FNN and MCMit-T do not fit on the QubiC ZCU216 RFSoC at all, every deployable model meets the sub-$\mu$s latency budget ($\leq74$\,ns), and KLiNQ (despite its small per-qubit parameter count) becomes the most resource-hungry deployable option ($\sim$70\% of LUTs/DSPs) once replicated across the array, whereas HERQULES and MCMit-CNN stay well within budget.

\takeaway{\textbf{Takeaway \#4: The smallest capable models win.} Two compact discriminators, MCMit-CNN and HERQULES, \textit{match or beat} models a thousand times larger, so on a limited-capacity FPGA controller, there is no tradeoff to negotiate: the most accurate \emph{deployable} discriminator is also among the smallest. What a model does with its input matters far more than how large it is.}

\subsection{Where Do Qubit-state Discriminators Fail?}
\label{subsec:failures}

\begin{wrapfigure}{r}{0.4\textwidth}
\centering
\includegraphics[width=0.4\textwidth]{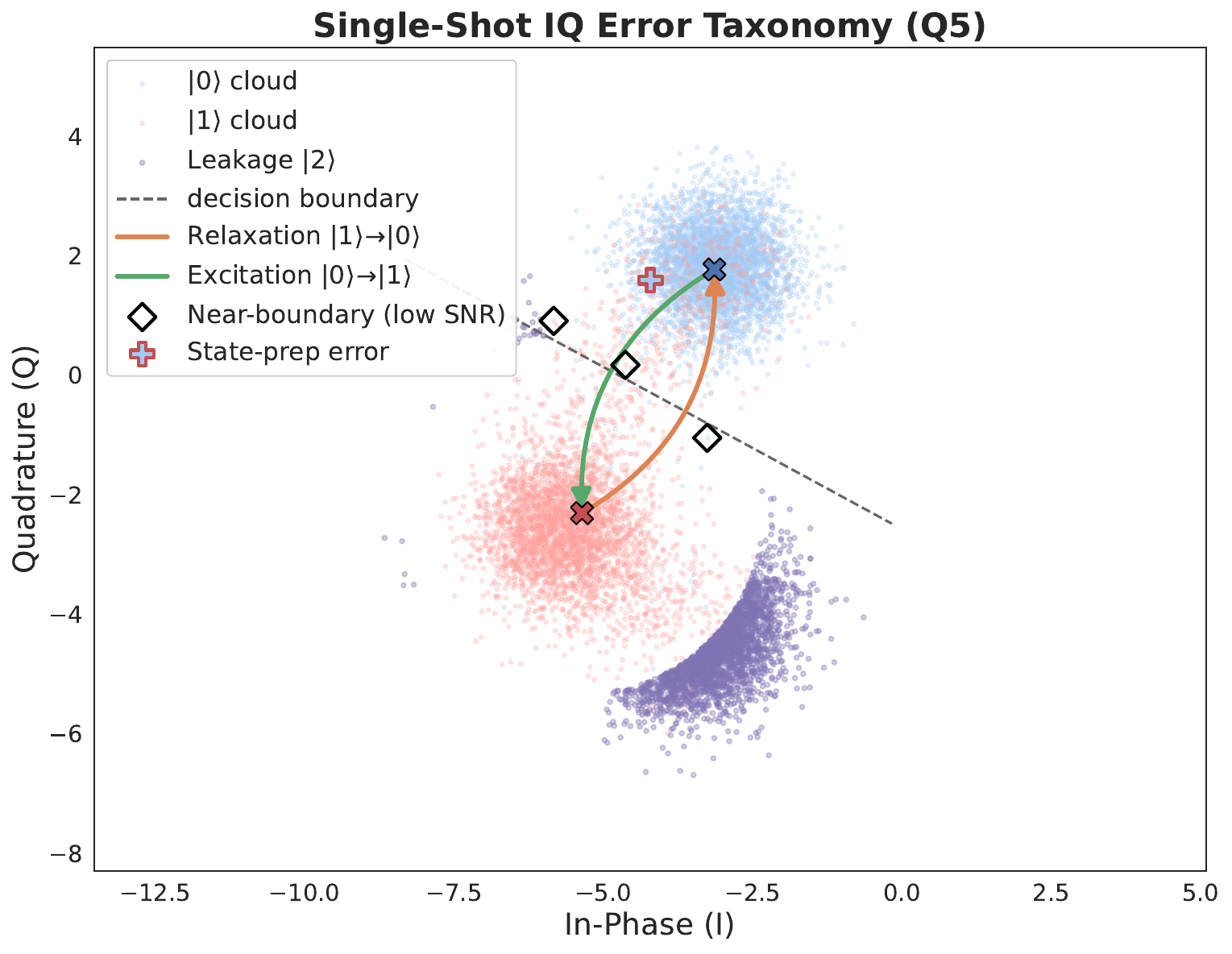}
\caption{Single-shot error taxonomy in the integrated IQ plane (Q5): 
Near-boundary ambiguity is intrinsic to the cloud overlap and dominates in practice. We provide detailed error taxonomy in App.~\ref{app:error-modes}
}
\label{fig:iq_taxonomy}
\end{wrapfigure}

Knowing how accurate discriminators are (\S\ref{subsec:discriminator_perforamnce}) is only half the picture; equally important is \emph{where} they fail. Misclassified shots are not spread uniformly: certain physical error types systematically confuse even the best models, and the difficulty can depend on the trace length.

\myparagraph{What counts as a failure}
A \emph{failure} is a shot assigned to the wrong state. In the integrated IQ plane each prepared state forms a Gaussian cloud split by the decision boundary (Fig.~\ref{fig:iq_taxonomy}); misclassifications come from mid-trace relaxation/excitation, excitation to $|2\rangle$, state-prep errors, and, most commonly, \emph{near-boundary, low-SNR} shots in the cloud overlap. We define the four modes and the attribution procedure in App.~\ref{app:error-modes}.

\myparagraph{Research question and hypothesis}
\rqbox{red}{\textbf{RQ\#5:} \textit{Which readout errors are the hardest to discriminate and why?}}

We hypothesize that mid-measurement relaxation and excitation produce the hardest traces, and that this difficulty grows with readout length.

\myparagraph{Methodology}
We attribute every misclassified shot to one of three modes (\textit{near-boundary}, \textit{mid-readout $T_1$ decay (relaxation)}, and \textit{leakage/excitation}; App.~\ref{app:error-modes}) and report per-mode rates for two high-accuracy discriminators, MCMit-CNN and HERQULES, on a mid-SNR (Q3) and the best (Q5) qubit at $1\,\mu$s and $400$\,ns, plus the full five-qubit MCMit-CNN breakdown (Table~\ref{tab:error_modes}).

\begin{figure}[t]
\centering
\includegraphics[width=0.85\linewidth]{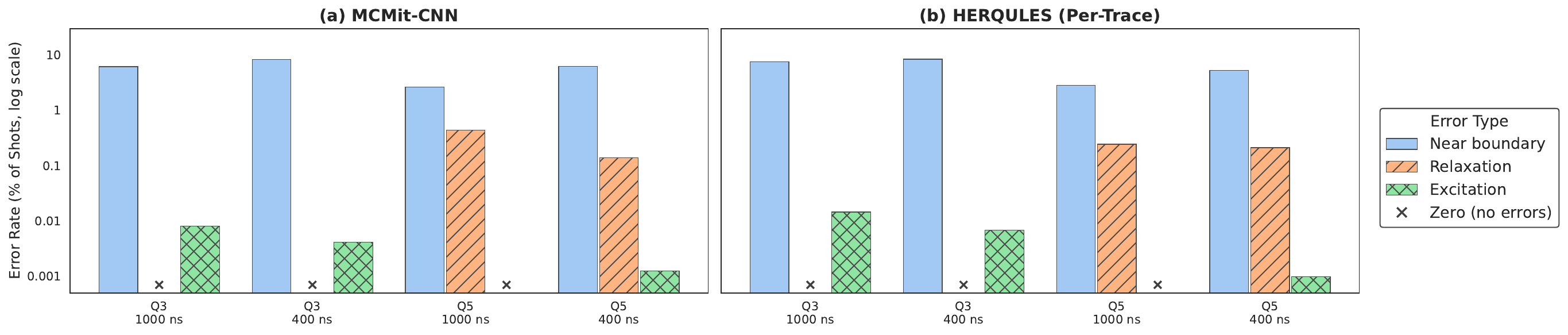}
\caption{Per-mode error breakdown (\% of shots, log scale) for HERQULES and MCMit-CNN on Q3 and Q5 at the $1\,\mu$s and $400$\,ns windows (\S~\ref{subsec:failures}). \textit{Near-boundary low-SNR ambiguity (blue) dominates everywhere.}}
\label{fig:error_breakdown}
\end{figure}

\myparagraph{Results}
Fig.~\ref{fig:error_breakdown} decomposes every misclassified shot. One mode dominates everywhere: \textit{near-boundary, low-SNR ambiguity} accounts for essentially all error on the mid-SNR qubit ($99.9\%$/$\sim\!100\%$ of MCMit-CNN's Q3 errors at $1000$/$400$\,ns) and the majority on the best qubit ($85.7\%$ at $1000$\,ns, $97.8\%$ at $400$\,ns). Excitation errors stay at $10^{-3}$--$10^{-2}\%$ of shots. The second, \textit{mid-readout $T_1$ decay/relaxation}, concentrates on the high-SNR qubit: $14.3\%$ of MCMit-CNN's Q5 errors at $1000$\,ns, shrinking to $2.2\%$ at $400$\,ns as the shorter window gives relaxation less time to strike; on Q3, the boundary floor is high enough that $T_1$ never surfaces.

\begin{table}[t]
\centering
\scriptsize
\setlength{\tabcolsep}{4pt}
\caption{MCMit-CNN per-qubit error-mode breakdown at $1\,\mu$s: each mode as a share of that qubit's \emph{errors}.}
\label{tab:error_modes}
\begin{tabular}{l r rrr}
\toprule
Qubit & Error rate & Near-boundary & Relaxation ($T_1$) & Leakage \\
\midrule
Q1 & $3.2\%$ & $100.0\%$ & $0.0\%$ & $0.0\%$ \\
Q2 (hardware) & $25.8\%$ & $97.5\%$ & $2.3\%$ & $0.2\%$  \\
Q3 & $6.1\%$ & $99.9\%$ & $0.0\%$ & $0.1\%$  \\
Q4 & $5.6\%$ & $99.4\%$ & $0.5\%$ & $0.04\%$  \\
Q5 & $3.0\%$ & $85.7\%$ & $14.3\%$ & $0.0\%$ \\
\midrule
All qubits & --- & $\mathbf{97.4\%}$ & $2.4\%$ & $0.2\%$ \\
\bottomrule
\end{tabular}
\end{table}

\myparagraph{Analysis}
Table~\ref{tab:error_modes} extends this to the whole register: pooled over five qubits, near-boundary ambiguity dominates ($97.4\%$), and $T_1$ relaxation is the only other appreciable mode ($2.4\%$); only the cleanest qubit (Q5) shows a visible $T_1$ dent. Mid-trace relaxation \emph{is} a real mode (its $T_1$ share on Q5 grows with window length as predicted), but it is not dominant. The two are layered, not competing: boundary ambiguity saturates the harder qubits so completely that no other mode shows, and only once a qubit clears it does the $T_1$ floor appear. This is why the two best models fail in the \emph{same} proportions despite a $2\times$ rate gap; both walls are physical (resonator SNR, qubit relaxation) and neither moves with architecture.

\takeaway{\textbf{Takeaway \#5: We have hit physics, not algorithms.} Some $99.8\%$ of every surviving error is device-set (low-SNR boundary ambiguity ($97.4\%$) plus mid-trace $T_1$ decay ($2.4\%$)), leaving nothing for a smarter classifier to recover. Discrimination is effectively solved; further fidelity must come from better readout hardware and longer-lived qubits, not a better model.}

\section{Qubit-State Readout Impact on QEC Performance}
\label{sec:qec_impact}

We first ask how discriminator accuracy and measurement duration each affect the logical error rate (\S\ref{subsec:readout_length_qec}), then characterize the measurement-duration effect across all six codes on current hardware (\S\ref{subsec:regimes_qec}), and finally on projected future hardware (\S\ref{subsec:regimes_qec_fixed}).

\subsection{How Does Discriminator Accuracy Impact Logical Error Rates?}
\label{subsec:readout_length_qec}

We start by isolating the discriminator's own contribution to the logical error rate.

\myparagraph{Research question and hypothesis}
\rqbox{green}{\textbf{RQ\#6:} \textit{How does discriminator accuracy impact LERs?}}

We hypothesize that a more accurate discriminator does lower the LER (the accuracy ranking of \S\ref{sec:discriminators} should carry over) but \emph{less} than measurement duration does, because a discriminator's accuracy swings far more across measurement durations than across models at a fixed length (\S\ref{sec:discriminators}). We further expect the effect to surface only at small code distances, where the LER is responsive; at large distances under current noise, the LER saturates, and the discriminator becomes irrelevant.

\myparagraph{Methodology} We first establish a faithful simulator, then use it to compare discriminators. \emph{Simulator:} We simulate a surface-code lattice-surgery logical CNOT between two patches, decoded with minimum-weight perfect matching, where a shot fails if any of the three logical observables is recovered incorrectly, in two independent circuit-level tools, ECCentric~\cite{Wierkowska2025Eccentric} and lattice-sim~\cite{Maurya2025Synchronization}. Both use the SOTA IBM Heron~r3 (\texttt{ibm\_boston}) noise model\footnote{CZ error $1.191\times10^{-3}$, SX error $1.637\times10^{-4}$, $T_1=284.95\,\mu$s, $T_2=322.68\,\mu$s.} and the length-dependent MCMit-CNN readout (the strongest deployable discriminator, \S\ref{sec:discriminators}); the swept length sets \emph{both} this flip error \emph{and} the measurement duration, so fidelity and idle $T_1/T_2$ decoherence trade off together, as on hardware. We sweep $d\in\{5,7,9,11,13\}$. The two simulators diverge on the surface-code logical CNOT (App.~\ref{app:sim-comparison}): ECCentric has the LER saturated near $0.5$ at every distance because the circuit is above threshold, while lattice-sim's narrower released noise-model scope yields a pronounced U-shaped optimum orders of magnitude lower. Given its broader noise-model coverage, we adopt \textbf{ECCentric} for the discriminator comparisons that follow. \emph{Discriminators:} having selected ECCentric, we then feed each of the seven discriminators of \S\ref{sec:discriminators} (its own per-qubit flip-error curve) into the simulator across four codes (surface, color, heavy-hex, Bacon-Shor) in \emph{memory}, at a responsive distance ($d=5$) and a saturated one ($d=12$).

\begin{figure}[t]
\centering
\includegraphics[width=0.85\linewidth]{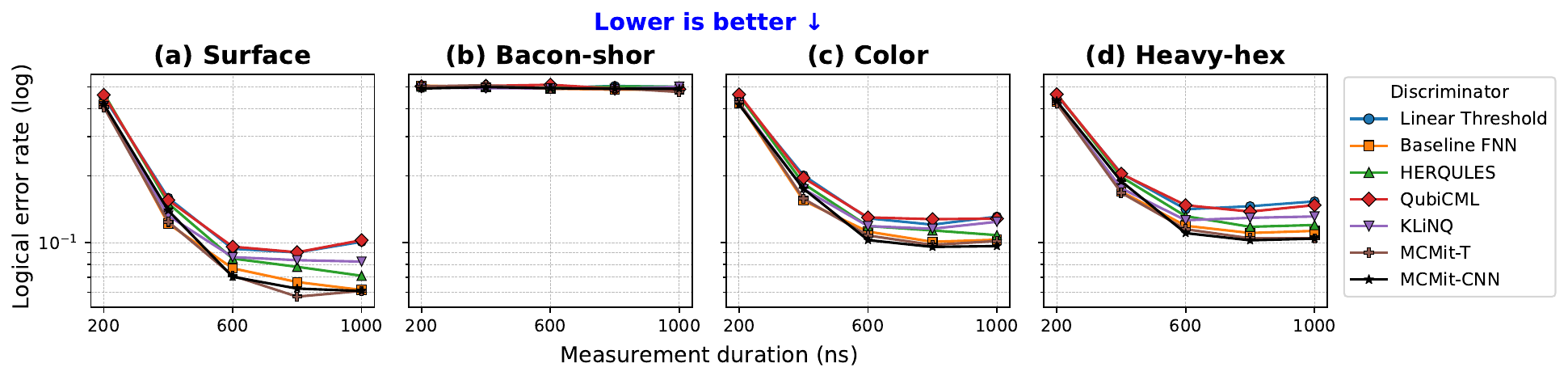}
\vspace{-3pt}
\caption{LER vs.\ measurement duration at $d=5$ (current noise), across four codes (\S\ref{subsec:readout_length_qec}). \textit{Measurement duration swings the LER by up to $\sim$590\% (surface code); at a fixed length, discriminator choice spreads it by $\sim$10--39\% (up to $69\%$). The best discriminator is always an MCMit variant, the worst QubiCML or the Linear threshold.}}
\label{fig:ler_discriminators_d5}
\vspace{-3pt}
\end{figure}

\myparagraph{Results} Fig.~\ref{fig:ler_discriminators_d5} sweeps all discriminators across four codes at $d=5$, exposing two effects at once. \emph{(1) measurement duration dominates}: along the MCMit-CNN line the LER swings by up to $\sim$590\% from $200$ to $1000$\,ns (surface $0.419$ to $0.061$; $\sim$415\% averaged over the three responsive codes). \emph{(2) Discriminator choice matters, second}: at a fixed length the spread across discriminators grows with the window, from $\sim$10\% at $200$\,ns to $\sim$39\% at $1000$\,ns, peaking at $69\%$ on surface at $1000$\,ns (QubiCML $0.103$ vs.\ MCMit-CNN $0.061$), and averages $\sim$26\%. The ordering is consistent: the best discriminator is always an MCMit variant and the worst is QubiCML or the Linear threshold, exactly the per-shot accuracy ranking of \S\ref{sec:discriminators}. Bacon-Shor is saturated ($\sim$0.5) and insensitive to either knob.

\begin{figure}[t]
\centering
\includegraphics[width=0.85\linewidth]{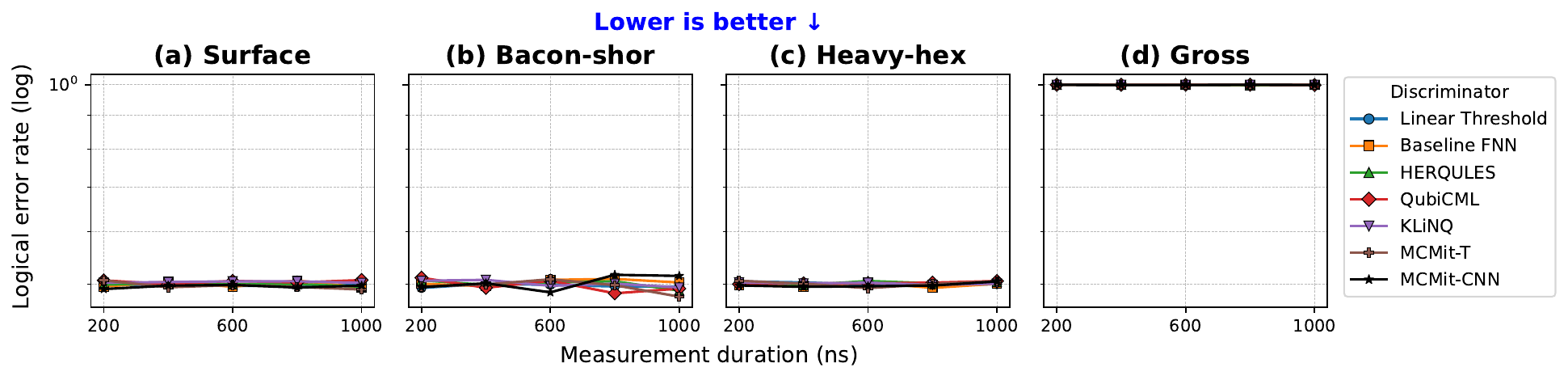}
\vspace{-3pt}
\caption{LER vs.\ measurement duration at $d=12$ (current noise), one line per discriminator, across four codes (\S\ref{subsec:readout_length_qec}). \textit{Both effects vanish: the measurement-duration gap is $\leq6\%$ and the discriminator spread $\sim$2.4\% (top $7.5\%$). All codes are saturated (surface/Bacon/heavy-hex $\approx0.5$, Gross $\approx1.0$), so the ranking flips randomly.}}
\label{fig:ler_discriminators_d12}
\vspace{-3pt}
\end{figure}

\myparagraph{Analysis} Both effects are confined to small, responsive distances. Repeating the four-code sweep at $d=12$ (Fig.~\ref{fig:ler_discriminators_d12}) flattens everything: the measurement-duration gap along the MCMit-CNN line collapses to $\leq6\%$ (surface $1\%$, heavy-hex $2\%$, Bacon-Shor $6\%$, Gross $0\%$), and the discriminator spread shrinks to $\sim$2.4\% on average (top $7.5\%$). At this distance, the codes are saturated (surface/Bacon/heavy-hex hover near $0.5$ from even-distance degeneracy, and Gross near $1.0$), so the discriminator ranking flips randomly from one measurement duration to the next: these residual gaps are sampling noise, not a fidelity signal. This confirms the hypothesis: under current noise, both knobs act only where the LER is responsive (small $d$); once scaling pushes the code deep above threshold, neither measurement duration nor discriminator choice moves the logical error rate.

\takeaway{\textbf{Takeaway \#6: Measurement duration leads, the discriminator follows.} At small distances, measurement duration is the dominant knob (up to $\sim$590\% LER swing), yet the discriminator still plays a clear role (up to $69\%$), carrying the accuracy ranking of \S\ref{sec:discriminators} to the QEC level. At large distances, neither matters: both effects collapse to noise ($\leq6\%$ and $\sim$2.4\% at $d=12$).}

\subsection{The Measurement-Duration Effect Across Codes and Noise Models}
\label{subsec:regimes_qec}

Section~\ref{subsec:readout_length_qec} isolated the discriminator-vs-duration comparison on four codes at two distances. Fixing the strongest discriminator (MCMit-CNN), we now characterize the readout-\emph{length} effect itself on \textbf{current and near-term hardware}, across all six codes (\S\ref{overview:taxonomies}) and the full distance range $d\in\{5,\dots,13\}$, under three \texttt{ibm\_boston}-based noise models. 

\myparagraph{Research question and hypothesis}
\rqbox{green}{\textbf{RQ\#7:} \textit{On current hardware, how does measurement duration affect the logical error rate, and how does hardware quality change the payoff?}}

We hypothesize that the measurement-duration effect (a steep LER drop into a $\sim$600\,ns plateau) appears only at small distances and is driven by the \emph{short-trace flip error}, so it vanishes once the flip is removed, and that the payoff \emph{grows} on poorer-coherence, backlog-limited hardware, where a long window costs both idle decoherence and decoding latency.

\myparagraph{Methodology} We sweep the readout length from $200$ to $1000$\,ns on all six codes with ECCentric and the MCMit-CNN readout, over $d\in\{5,7,9,11,13\}$, under three \texttt{ibm\_boston}-based models: current noise, a \emph{decoherence-only} model (readout flip removed), and a \emph{low-coherence} model ($T_1/T_2{=}190/130\,\mu$s) with a duration-independent readout and a decoder-\emph{backlog} penalty.

\begin{figure}[t]
\centering
\includegraphics[width=0.85\linewidth]{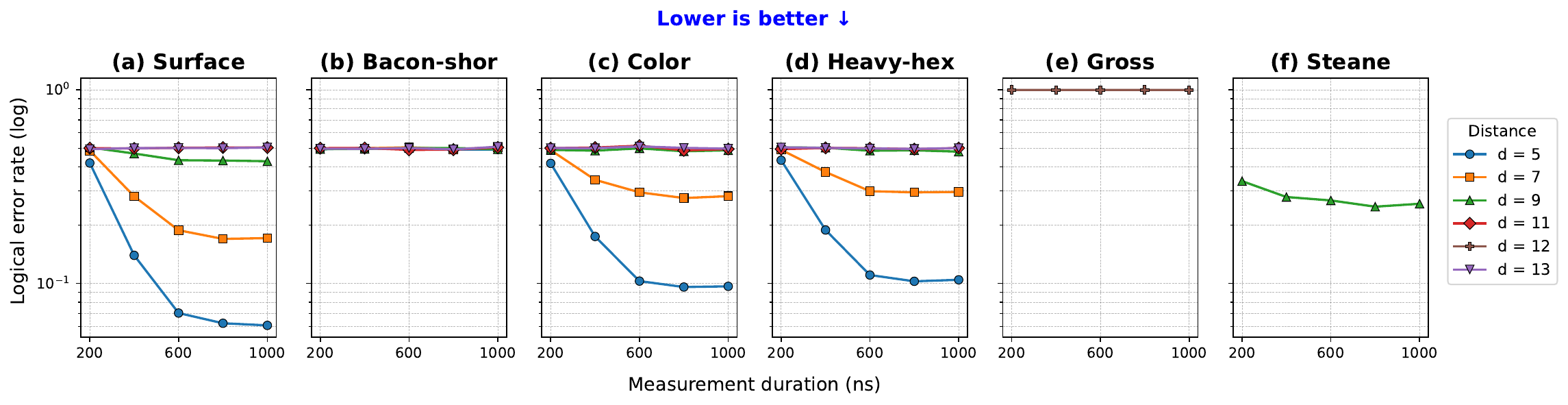}
\vspace{-3pt}
\caption{LER vs.\ measurement duration, six codes, \textbf{current} \texttt{ibm\_boston} noise. \textit{Measurement duration matters only at small $d$, with a $\sim$600\,ns plateau; larger $d$ saturates near $0.5$ and Bacon-Shor/Gross are dead.}}
\label{fig:six_current}
\vspace{-3pt}
\end{figure}

\myparagraph{Results} Fig.~\ref{fig:six_current} confirms the \S\ref{subsec:readout_length_qec} picture across all six codes: measurement duration drives the LER only at small distances, where surface, color, and heavy-hex fall sharply from $200$ to $\sim$600\,ns at $d{=}5$--$7$ and then plateau, while larger distances saturate near $0.5$ and the measurement-limited Bacon-Shor and gate-limited Gross codes are dead at every length. The $\sim$600\,ns plateau is therefore a general, code-independent feature of the responsive regime, not a surface-code artifact. Removing the readout flip entirely confirms it is the sole cause of the optimum, and reveals that Bacon-Shor's LER collapses to $\approx0$ without it (App.~\ref{app:decoherence-only}).

\myparagraph{Analysis} The payoff \emph{grows} on worse hardware. Here, readout fidelity is held fixed (duration-independent), so there is no short-trace optimum (shorter is strictly better), and with $T_1/T_2$ roughly halved plus a decoder-backlog penalty, a longer window costs twice: more idle decoherence \emph{and} more backlog. Shortening then cuts the LER by up to $\sim$27\%, so measurement duration matters \emph{more}, not less, on poorer, latency-constrained hardware.

\begin{figure}[t]
\centering
\includegraphics[width=0.85\linewidth]{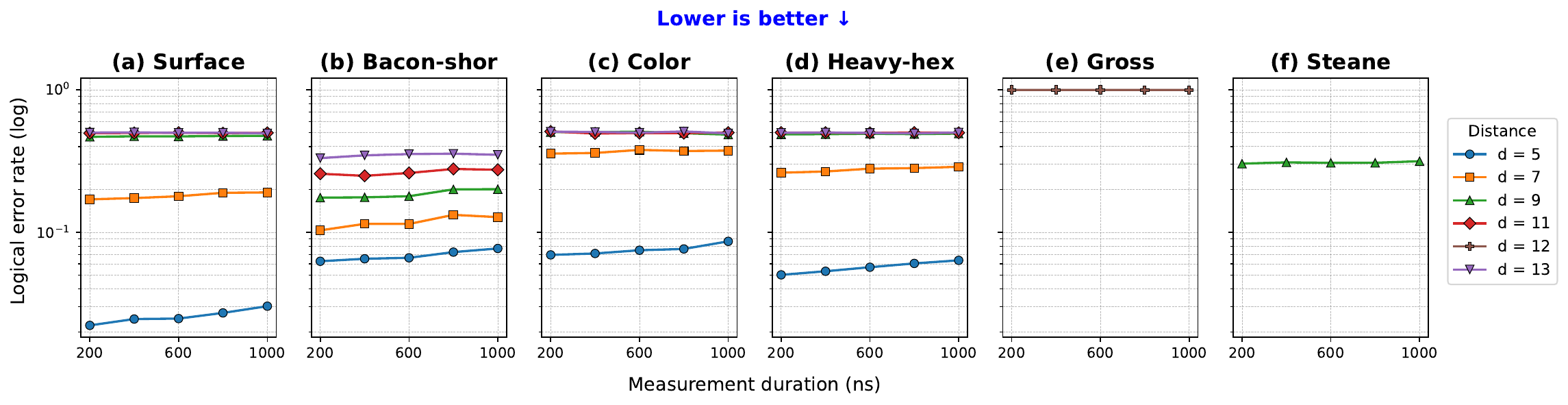}
\vspace{-3pt}
\caption{LER vs.\ measurement duration, six codes, \textbf{low-coherence} model ($T_1/T_2{=}190/130\,\mu$s, decoder \textbf{backlog} on). \textit{A longer window costs both idle decoherence and decoding backlog, so shortening cuts the LER by up to $\sim$27\%; measurement duration matters more on poorer, latency-constrained hardware.}}
\label{fig:six_backlog}
\vspace{-3pt}
\end{figure}

\takeaway{\textbf{Takeaway \#7: The readout error sets the optimum.} The $\sim$600\,ns optimum is caused \emph{entirely} by the short-trace readout error, and is strongest on measurement-limited codes. On poorer-coherence, backlog-limited hardware, shortening helps even more (up to $\sim$27\%), since a long window costs both idle decoherence and decoding latency.}

\subsection{Measurement Duration on Future Hardware}
\label{subsec:regimes_qec_fixed}

\begin{wrapfigure}{r}{0.45\textwidth}
\centering
\resizebox{\linewidth}{!}{%
\begin{tikzpicture}[font=\scriptsize,
  cell/.style={rounded corners=2pt,minimum width=2.7cm,minimum height=1.0cm,align=center,inner sep=2pt,draw},
  matters/.style={cell,draw=blue!60!black,fill=blue!14},
  free/.style={cell,draw=green!55!black,fill=green!12},
  flat/.style={cell,draw=gray!55,fill=gray!12}]
  \node[align=center] at (0,1.5) {\textbf{fixed-}\\\textbf{fidelity}};
  \node[align=center] at (3,1.5) {\textbf{length-dep.}\\\textbf{readout}};
  \node[align=right,anchor=east] at (-1.5,0.6) {\textbf{low LER}\\(responsive)};
  \node[align=right,anchor=east] at (-1.5,-0.6) {\textbf{saturated}\\(LER$\approx$0.5)};
  \node[free] at (0,0.6) {\textbf{free}\\shorter is better};
  \node[matters] at (3,0.6) {\textbf{first-class}\\$\sim$600\,ns optimum};
  \node[flat,minimum width=5.7cm] at (1.5,-0.6) {\textbf{flat}: measurement duration irrelevant};
\end{tikzpicture}}
\caption{Measurement duration is a first-class QEC knob in only one regime: a low-LER regime \emph{and} a length-dependent readout. Elsewhere, it is free or irrelevant.}
\label{fig:regime_map}
\end{wrapfigure}
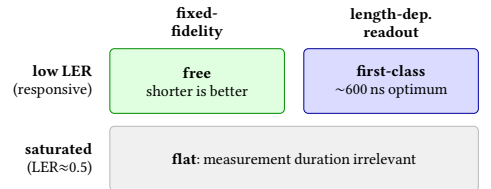

We now ask how the measurement duration impact shifts as hardware improves. Fig.~\ref{fig:regime_map} shows that measurement duration is a first-class knob in one regime, a low-LER operating point with a length-dependent readout, and lowering the error rate is what moves codes into it.

\myparagraph{Research question and hypothesis}
\rqbox{green}{\textbf{RQ\#8:} \textit{As hardware improves, does measurement duration become a more important QEC knob?}}

We hypothesize that improving readout fidelity \emph{alone} primarily benefits measurement-limited codes but leaves the system gate-limited, whereas lowering every error rate (a futuristic projection) lowers the LER into the responsive regime, where the $\sim$600\,ns optimum reappears across codes and distances.

\myparagraph{Methodology} On the same six-code memory sweep with ECCentric and the MCMit-CNN readout, we evaluate two projected-hardware models: a \emph{high-fidelity readout} model (readout error $/10$, gates and coherence at current \texttt{ibm\_boston}) and a \emph{futuristic} model (all errors $/10$, $T_1/T_2$ tripled).

\begin{figure}[t]
\centering
\includegraphics[width=0.85\linewidth]{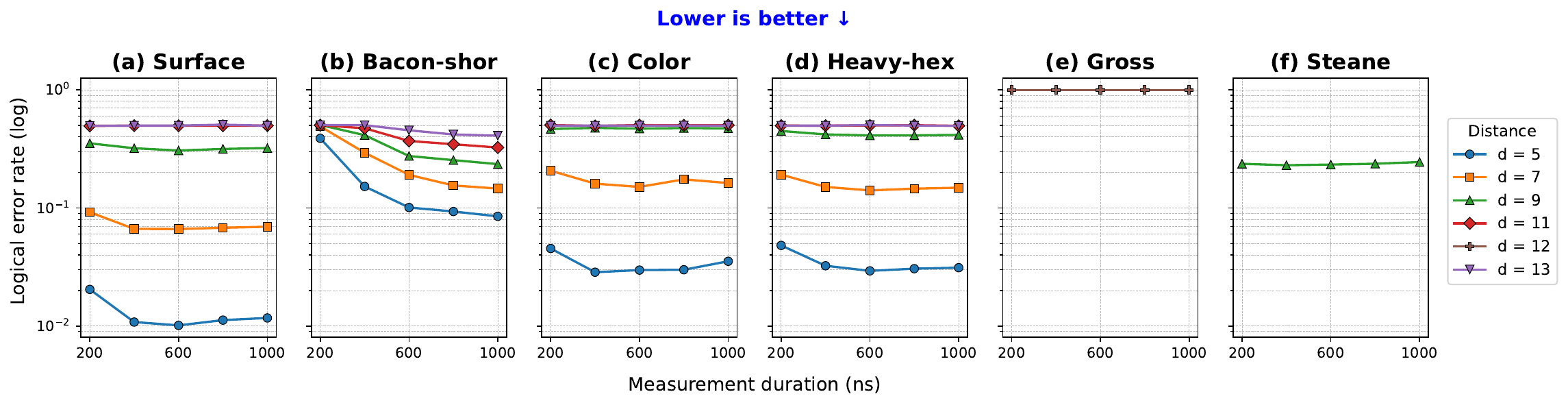}
\vspace{-3pt}
\caption{LER vs.\ measurement duration, six codes, \textbf{high-fidelity readout}. \textit{Better readout alone lowers small-$d$ LER and revives Bacon-Shor, but higher $d$ stays saturated, so the system is gate-limited.}}
\label{fig:six_highfid}
\vspace{-3pt}
\end{figure}

\myparagraph{Results} Improving readout fidelity tenfold (Fig.~\ref{fig:six_highfid}) lowers the small-$d$ LER and, most visibly, revives the measurement-limited Bacon-Shor, but the higher distances stay saturated, and the LER still rises with $d$: the system remains \emph{gate-limited}, so better readout alone widens the responsive regime only partway.

\begin{figure}[t]
\centering
\includegraphics[width=0.85\linewidth]{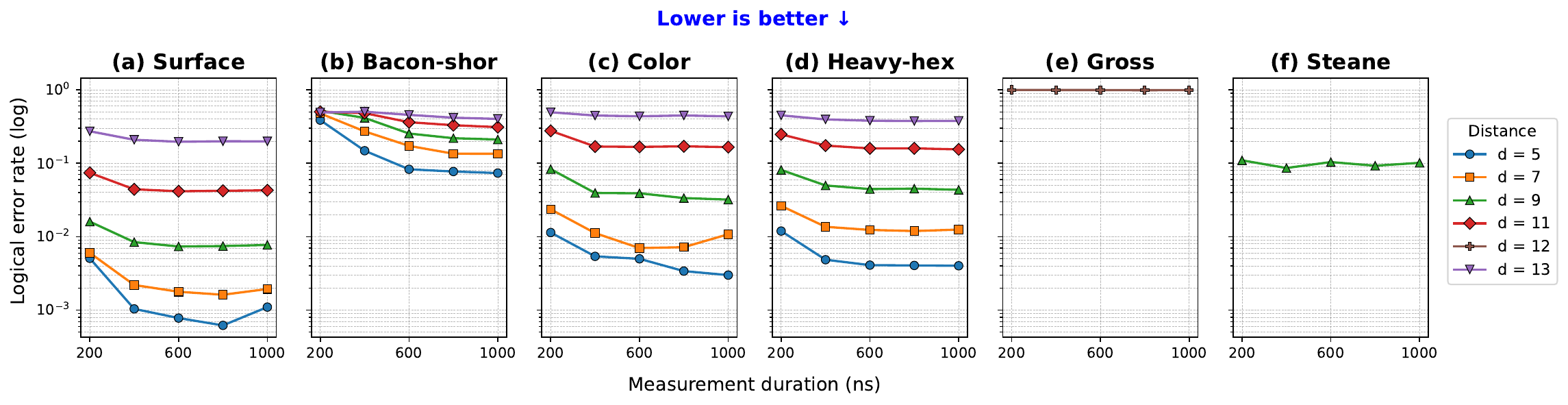}
\vspace{-3pt}
\caption{LER vs.\ measurement duration, six codes, \textbf{futuristic} noise. \textit{The lower prefactor widens the responsive regime: a $\sim$600--800\,ns optimum appears across codes and Bacon-Shor comes alive; LER still rises with $d$.}}
\label{fig:six_fut_herqules}
\vspace{-3pt}
\end{figure}

\myparagraph{Analysis} Scaling all error rates down (futuristic; Fig.~\ref{fig:six_fut_herqules}) does move the system into the responsive regime: a clear $\sim$600--800\,ns optimum reappears across surface, color, and heavy-hex, more distances respond, and Bacon-Shor tracks measurement duration strongly. The LER still rises with distance (the $/10$ scaling lowers the prefactor without crossing the threshold; App.~\ref{app:futuristic}), so measurement duration becomes a first-class knob for more codes and distances as hardware improves, but reaching genuine fault tolerance requires the gates to improve too, not just readout alone.

\takeaway{\textbf{Takeaway \#8: Better hardware makes measurement duration matter more.} Better readout alone mainly revives measurement-limited codes; scaling every error rate widens the responsive regime so the $\sim$600\,ns optimum reappears across codes and distances. Readout strategy thus matters more on future hardware, but only alongside better gates.}

\section{Related Work}

\myparagraph{Qubit-state discrimination}
ML discriminators have largely replaced matched-filter and threshold methods, from early SVM/NN trajectory classifiers~\cite{Magesan2015Machine} to deep networks for frequency-multiplexed readout~\cite{Lienhard2022Deep}, the matched-filter/NN HERQULES pipeline~\cite{Maurya2023Scaling}, real-time FPGA discrimination enabling mid-circuit measurement~\cite{Vora2024Ml}, end-to-end QICK/hls4ml workflows~\cite{Guglielmo2025End}, distillation-compressed networks~\cite{Guo2025Klinq}, and qudit-scale architectures~\cite{Mude2025Efficient}; on the physics side, prior work characterized the limits of fast dispersive readout~\cite{Walter2017Rapid,Heinsoo2018Rapid,Chen2023Transmon}. Each, however, evaluates a single discriminator under a fixed condition and reports physical-level accuracy in isolation, without a unified benchmark across discriminators, the accuracy/latency/FPGA-resource Pareto frontier, robustness across noise regimes, or a link to downstream QEC. We close all three gaps.

\myparagraph{QEC simulation and benchmarking}
\label{subsec:related_qec}
QEC theory and simulation are well developed: surface-code thresholds, overhead, and decoding~\cite{Fowler2012Surface,Dennis2002Topological,Gidney2021How}, fast stabilizer simulators~\cite{Gidney2021Stim}, MWPM decoders~\cite{Roffe2020Decoding}, real-time decoding~\cite{Battistel2023Real}, and high-threshold codes~\cite{Bravyi2024High}. Empirical benchmarking is sparser: small-code protocols on near-term hardware~\cite{Wootton2020Benchmarking,Finsterhoelzl2022Benchmarking}, the cross-simulator ECCentric study~\cite{Wierkowska2025Eccentric}, and code-comparison parameters, decoder, and controller benchmarks~\cite{Chatterjee2024Magic,Zhao2024Benchmarking,Harper2025Characterising,Kurman2024Benchmarking}. All treat the measurement error rate as fixed, not set by the readout hardware and discriminator. We instead derive per-shot measurement errors from real IQ traces under varying lengths and discriminators, injecting them into QEC simulation to link readout configuration to logical error rate.

\myparagraph{Quantum computing benchmarking}
This targets gates, circuits, and compilers: error characterization on IBM devices, which identifies measurement as the dominant source~\cite{Patel2020Experimental}, application- and low-level benchmark suites~\cite{Tomesh2022Supermarq,Li2022Qasmbench}, benchmarking methodology and scalable processor/gate protocols~\cite{Lorenz2025Systematic,Mckay2023Benchmarking,Proctor2024Benchmarking}, compiler benchmarks~\cite{Nation2025Benchmarking}, systems-level resource management and virtualization~\cite{Giortamis2025Qos,Giortamis2025Qonductor,Tornow2025Qvm}, and FPGA control and feedback~\cite{Xu2021Qubic,Tian2025Artery}. None treats the readout pipeline as a first-class, cross-layer component spanning the signal, discriminator, and QEC layers.

\section{Concluding Remarks and Discussion}

We present \projectname, the first end-to-end benchmarking framework and empirical study of how qubit-state readout shapes QEC performance, treating readout as a \emph{systems-level} component of the quantum stack rather than a niche calibration step. Using experimentally-extracted readout traces, we trace this relationship across three coupled layers: the signal level, the discriminator level (and its FPGA cost), and the QEC level (logical error rates across six codes and hardware regimes). The throughline is that measurement duration, discriminator architecture, and QEC code behavior are \emph{coupled} choices, to be reasoned about together rather than optimized in isolation, and that their impact on the logical error rate is conditional rather than universal.

We distill the higher-level lessons, each grounded in the takeaways above: which
knobs matter most, which can be deprioritized, and where the limits lie.

\begin{itemize}
\item \textbf{Readout is a systems-level decision, not a calibration step.} The readout window couples the signal, the discriminator, and the QEC cycle, so tuning it in isolation leaves both latency and logical-error rate on the table (\textbf{Takeaways~\#1--\#8}).

\item \textbf{Cut measurement duration first: it is the dominant, nearly-free lever.} 
Fidelity saturates well before the typical $1\,\mu$s window, so readout can be cut to $\sim$600\,ns at negligible per-shot cost (right where the logical error rate also plateaus), reclaiming $\sim$40\% of readout latency and shortening every QEC cycle (\textbf{Takeaways~\#1,~\#2,~\#3,~\#7}).




\item \textbf{Discrimination is physically, not algorithmically, limited.} Roughly $99.8\%$ of residual misclassifications are device-set (low-SNR boundary ambiguity and mid-trace $T_1$ decay), which no architecture or window length removes. Further fidelity must come from better hardware (higher-SNR chains, longer-lived qubits), not cleverer models (\textbf{Takeaway~\#5}).

\item \textbf{Pick the cheapest capable discriminator.} Since the model cannot move that physical wall and its logical-error impact is bounded anyway, there is no reason to pay for a large one. The best deployable discriminator is also among the smallest, freeing FPGA logic
and latency for the rest of the control stack \textbf{Takeaways~\#4,~\#6}).

\item \textbf{Readout length pays off only in specific regimes, widening with hardware.} LER improves only where the LER is responsive, and readout error is length-dependent: at accessible code distances, for measurement-limited codes (e.g.\ Bacon-Shor), and on low-coherence hardware. This regime is narrow today but broadens as gates and coherence improve, making readout a more important QEC knob on future hardware \textbf{Takeaways~\#7,~\#8}).
\end{itemize}

\section*{Acknowledgments}
This work is funded by the Bavarian State Ministry of Science and the Arts as part of the Munich Quantum Valley (MQV), grant number 6090181. B.L. acknowledges support from the German Federal Ministry of Research, Technology, and Space (BMFTR) through the program EQuIPS (Grant No. 13N17232).

\bibliographystyle{ACM-Reference-Format}
\bibliography{references}

@article{Terhal2015Quantum,
  title = {Quantum error correction for quantum memories},
  author = {Terhal, Barbara M.},
  journal = {Rev. Mod. Phys.},
  volume = {87},
  issue = {2},
  pages = {307--346},
  numpages = {40},
  year = {2015},
  month = {Apr},
  publisher = {American Physical Society},
  doi = {10.1103/RevModPhys.87.307},
  url = {https://link.aps.org/doi/10.1103/RevModPhys.87.307}
}

@article{Yang2022Fpga,
    author = {Yang, Yuchen and Shen, Zhongtao and Zhu, Xing and Wang, Ziqi and Zhang, Gengyan and Zhou, Jingwei and Jiang, Xun and Deng, Chunqing and Liu, Shubin},
    title = {FPGA-based electronic system for the control and readout of superconducting quantum processors},
    journal = {Review of Scientific Instruments},
    volume = {93},
    number = {7},
    pages = {074701},
    year = {2022},
    month = {07},
    issn = {0034-6748},
    doi = {10.1063/5.0085467},
    url = {https://doi.org/10.1063/5.0085467},
}

@article{Duarte2018Fast,
    author = "Duarte, Javier and others",
    title = "{Fast inference of deep neural networks in FPGAs for particle physics}",
    eprint = "1804.06913",
    archivePrefix = "arXiv",
    primaryClass = "physics.ins-det",
    reportNumber = "FERMILAB-PUB-18-089-E",
    doi = "10.1088/1748-0221/13/07/P07027",
    journal = "JINST",
    volume = "13",
    number = "07",
    pages = "P07027",
    year = "2018"
}

@inproceedings{Maurya2023Scaling,
author = {Maurya, Satvik and Mude, Chaithanya Naik and Oliver, William D. and Lienhard, Benjamin and Tannu, Swamit},
title = {Scaling Qubit Readout with Hardware Efficient Machine Learning Architectures},
year = {2023},
isbn = {9798400700958},
publisher = {Association for Computing Machinery},
address = {New York, NY, USA},
url = {https://doi.org/10.1145/3579371.3589042},
doi = {10.1145/3579371.3589042},
booktitle = {Proceedings of the 50th Annual International Symposium on Computer Architecture},
articleno = {7},
numpages = {13},
location = {Orlando, FL, USA},
series = {ISCA '23}
}

@article{Acharya2023Multiplexed,
  title={Multiplexed superconducting qubit control at millikelvin temperatures with a low-power cryo-CMOS multiplexer},
  author={Acharya, Rohith and Brebels, Steven and Grill, Alexander and Verjauw, Jeroen and Ivanov, Ts and Lozano, D Perez and Wan, Danny and Van Damme, Jacques and Vadiraj, AM and Mongillo, Massimo and others},
  journal={Nature Electronics},
  volume={6},
  number={11},
  pages={900--909},
  year={2023},
  publisher={Nature Publishing Group UK London}
}

@article{Knill1997Theory,
  title = {Theory of quantum error-correcting codes},
  author = {Knill, Emanuel and Laflamme, Raymond},
  journal = {Phys. Rev. A},
  volume = {55},
  issue = {2},
  pages = {900--911},
  numpages = {0},
  year = {1997},
  month = {Feb},
  publisher = {American Physical Society},
  doi = {10.1103/PhysRevA.55.900},
  url = {https://link.aps.org/doi/10.1103/PhysRevA.55.900}
}

@article{Wierkowska2025Eccentric,
author = {\'{S}wierkowska, Aleksandra and Pflieger, Jannik and Giortamis, Emmanouil and Bhatotia, Pramod},
title = {ECCentric: An Empirical Analysis of Quantum Error Correction Codes},
year = {2026},
issue_date = {June 2026},
publisher = {Association for Computing Machinery},
address = {New York, NY, USA},
volume = {10},
number = {2},
url = {https://doi.org/10.1145/3805635},
doi = {10.1145/3805635},
journal = {Proc. ACM Meas. Anal. Comput. Syst.},
month = may,
articleno = {37},
numpages = {33}
}

@inproceedings{Giortamis2025Qos,
author = {Giortamis, Emmanouil and Rom\~{a}o, Francisco and Tornow, Nathaniel and Bhatotia, Pramod},
title = {QOS: quantum operating system},
year = {2025},
isbn = {978-1-939133-47-2},
publisher = {USENIX Association},
address = {USA},
booktitle = {Proceedings of the 19th USENIX Conference on Operating Systems Design and Implementation},
articleno = {24},
numpages = {19},
location = {Boston, MA, USA},
series = {OSDI '25}
}

@inproceedings{Giortamis2025Qonductor,
author = {Giortamis, Emmanouil and Romao, Francisco and Tornow, Nathaniel and Lugovoy, Dmitry and Bhatotia, Pramod},
title = {Qonductor: A Cloud Orchestrator for Quantum Computing},
year = {2025},
isbn = {9798400714665},
publisher = {Association for Computing Machinery},
address = {New York, NY, USA},
url = {https://doi.org/10.1145/3712285.3759785},
doi = {10.1145/3712285.3759785},
booktitle = {Proceedings of the International Conference for High Performance Computing, Networking, Storage and Analysis},
pages = {728–745},
numpages = {18},
location = {
},
series = {SC '25}
}

@article{Tornow2025Qvm,
author = {Tornow, Nathaniel and Giortamis, Emmanouil and Bhatotia, Pramod},
title = {QVM: Quantum Gate Virtualization Machine},
year = {2025},
issue_date = {June 2025},
publisher = {Association for Computing Machinery},
address = {New York, NY, USA},
volume = {9},
number = {PLDI},
url = {https://doi.org/10.1145/3729290},
doi = {10.1145/3729290},
journal = {Proc. ACM Program. Lang.},
month = jun,
articleno = {187},
numpages = {26}
}

@inproceedings{Tian2025Artery,
author = {Tian, Wuwei and Lu, Liqiang and Tan, Siwei and Liang, Yun and Li, Tingting and Zhou, Kaiwen and Jia, Xinghui and Yin, Jianwei},
title = {ARTERY: Fast Quantum Feedback using Branch Prediction},
year = {2025},
isbn = {9798400712616},
publisher = {Association for Computing Machinery},
address = {New York, NY, USA},
url = {https://doi.org/10.1145/3695053.3731086},
doi = {10.1145/3695053.3731086},
booktitle = {Proceedings of the 52nd Annual International Symposium on Computer Architecture},
pages = {285–298},
numpages = {14},
location = {
},
series = {ISCA '25}
}

@inproceedings{Tannu2019Mitigating,
author = {Tannu, Swamit S. and Qureshi, Moinuddin K.},
title = {Mitigating Measurement Errors in Quantum Computers by Exploiting State-Dependent Bias},
year = {2019},
isbn = {9781450369381},
publisher = {Association for Computing Machinery},
address = {New York, NY, USA},
url = {https://doi.org/10.1145/3352460.3358265},
doi = {10.1145/3352460.3358265},
booktitle = {Proceedings of the 52nd Annual IEEE/ACM International Symposium on Microarchitecture},
pages = {279–290},
numpages = {12},
location = {Columbus, OH, USA},
series = {MICRO '52}
}

@article{Heinsoo2018Rapid,
  title = {Rapid High-fidelity Multiplexed Readout of Superconducting Qubits},
  author = {Heinsoo, Johannes and Andersen, Christian Kraglund and Remm, Ants and Krinner, Sebastian and Walter, Theodore and Salath\'e, Yves and Gasparinetti, Simone and Besse, Jean-Claude and Poto\ifmmode \check{c}\else \v{c}\fi{}nik, Anton and Wallraff, Andreas and Eichler, Christopher},
  journal = {Phys. Rev. Appl.},
  volume = {10},
  issue = {3},
  pages = {034040},
  numpages = {14},
  year = {2018},
  month = {Sep},
  publisher = {American Physical Society},
  doi = {10.1103/PhysRevApplied.10.034040},
  url = {https://link.aps.org/doi/10.1103/PhysRevApplied.10.034040}
}

@ARTICLE{Xu2021Qubic,

  author={Xu, Yilun and Huang, Gang and Balewski, Jan and Naik, Ravi and Morvan, Alexis and Mitchell, Bradley and Nowrouzi, Kasra and Santiago, David I. and Siddiqi, Irfan},

  journal={IEEE Transactions on Quantum Engineering}, 

  title={QubiC: An Open-Source FPGA-Based Control and Measurement System for Superconducting Quantum Information Processors}, 

  year={2021},

  volume={2},

  number={},

  pages={1-11},

  doi={10.1109/TQE.2021.3116540}}

@misc{IQM2024Garnet,
  title = {{IQM Garnet: 20-qubit Superconducting Quantum Computer}},
  author = {{IQM Quantum Computers}},
  year = {2024},
  howpublished = {\url{https://www.meetiqm.com/products/iqm-garnet}},
  note = {Accessed: 2026-06-17}
}

@misc{Rigetti2024Ankaa,
  title = {{Rigetti Computing Launches 84-Qubit Ankaa-3 System}},
  author = {{Rigetti Computing}},
  year = {2024},
  howpublished = {\url{https://investors.rigetti.com/news-releases/news-release-details/rigetti-computing-launches-84-qubit-ankaatm-3-system-achieves}},
  note = {Accessed: 2026-06-17}
}

@misc{UnknownXXXXIbma,  
  title = {IBM Quantum Cloud Compute Resources},
  key = {ibm-devices},
  publisher = {IBM Quantum},
  howpublished = {\url{https://quantum.ibm.com/services/resources}},
  note = {Accessed: 2025-20-08}
}

@misc{Vora2024Ml,
      title={ML-Powered FPGA-based Real-Time Quantum State Discrimination Enabling Mid-circuit Measurements}, 
      author={Neel R. Vora and Yilun Xu and Akel Hashim and Neelay Fruitwala and Ho Nam Nguyen and Haoran Liao and Jan Balewski and Abhi Rajagopala and Kasra Nowrouzi and Qing Ji and K. Birgitta Whaley and Irfan Siddiqi and Phuc Nguyen and Gang Huang},
      year={2024},
      eprint={2406.18807},
      archivePrefix={arXiv},
      primaryClass={quant-ph},
      url={https://arxiv.org/abs/2406.18807}, 
}

@article{Preskill2018Quantum,
  doi = {10.22331/q-2018-08-06-79},
  url = {https://doi.org/10.22331/q-2018-08-06-79},
  title = {Quantum {C}omputing in the {NISQ} era and beyond},
  author = {Preskill, John},
  journal = {{Quantum}},
  issn = {2521-327X},
  publisher = {{Verein zur F{\"{o}}rderung des Open Access Publizierens in den Quantenwissenschaften}},
  volume = {2},
  pages = {79},
  month = aug,
  year = {2018}
}

@misc{Guglielmo2025End,
      title={End-to-end workflow for machine learning-based qubit readout with QICK and hls4ml}, 
      author={Giuseppe Di Guglielmo and Botao Du and Javier Campos and Alexandra Boltasseva and Akash V. Dixit and Farah Fahim and Zhaxylyk Kudyshev and Santiago Lopez and Ruichao Ma and Gabriel N. Perdue and Nhan Tran and Omer Yesilyurt and Daniel Bowring},
      year={2025},
      eprint={2501.14663},
      archivePrefix={arXiv},
      primaryClass={quant-ph},
      url={https://arxiv.org/abs/2501.14663}, 
}

@misc{Guo2025Klinq,
      title={KLiNQ: Knowledge Distillation-Assisted Lightweight Neural Network for Qubit Readout on FPGA}, 
      author={Xiaorang Guo and Tigran Bunarjyan and Dai Liu and Benjamin Lienhard and Martin Schulz},
      year={2025},
      eprint={2503.03544},
      archivePrefix={arXiv},
      primaryClass={quant-ph},
      url={https://arxiv.org/abs/2503.03544}, 
}

@misc{Mude2025Efficient,
      title={Efficient and Scalable Architectures for Multi-Level Superconducting Qubit Readout}, 
      author={Chaithanya Naik Mude and Satvik Maurya and Benjamin Lienhard and Swamit Tannu},
      year={2025},
      eprint={2405.08982},
      archivePrefix={arXiv},
      primaryClass={quant-ph},
      url={https://arxiv.org/abs/2405.08982}, 
}

@INPROCEEDINGS{Lienhard2022Reinforcement,
       author = {{Lienhard}, Benjamin and {Vepsalainen}, Antti and {Hoffer}, Cole and {Govia}, Luke and {Andersen Woltz}, Vilhelm and {Kim}, David and {Melville}, Alexander and {Huffman}, Bethany and {Yoder}, Jonilyn and {Schwartz}, Mollie and {Orlando}, Terry and {Oliver}, William},
        title = "{Reinforcement Learning assisted Pulse Shaping for Superconducting Qubit Readout}",
    booktitle = {APS March Meeting Abstracts},
         year = 2022,
       series = {APS Meeting Abstracts},
       volume = {2022},
        month = mar,
          eid = {F41.002},
        pages = {F41.002},
       adsurl = {https://ui.adsabs.harvard.edu/abs/2022APS..MARF41002L}
}

@Article{Chen2023Transmon,
author={Chen, Liangyu
and Li, Hang-Xi
and Lu, Yong
and Warren, Christopher W.
and Kri{\v{z}}an, Christian J.
and Kosen, Sandoko
and Rommel, Marcus
and Ahmed, Shahnawaz
and Osman, Amr
and Bizn{\'a}rov{\'a}, Janka
and Fadavi Roudsari, Anita
and Lienhard, Benjamin
and Caputo, Marco
and Grigoras, Kestutis
and Gr{\"o}nberg, Leif
and Govenius, Joonas
and Kockum, Anton Frisk
and Delsing, Per
and Bylander, Jonas
and Tancredi, Giovanna},
title={Transmon qubit readout fidelity at the threshold for quantum error correction without a quantum-limited amplifier},
journal={npj Quantum Information},
year={2023},
month={Mar},
day={16},
volume={9},
number={1},
pages={26},
issn={2056-6387},
doi={10.1038/s41534-023-00689-6},
url={https://doi.org/10.1038/s41534-023-00689-6}
}

@article{Magesan2015Machine,
  title = {Machine Learning for Discriminating Quantum Measurement Trajectories and Improving Readout},
  author = {Magesan, Easwar and Gambetta, Jay M. and C\'orcoles, A. D. and Chow, Jerry M.},
  journal = {Phys. Rev. Lett.},
  volume = {114},
  issue = {20},
  pages = {200501},
  numpages = {5},
  year = {2015},
  month = {May},
  publisher = {American Physical Society},
  doi = {10.1103/PhysRevLett.114.200501},
  url = {https://link.aps.org/doi/10.1103/PhysRevLett.114.200501}
}

@article{Walter2017Rapid,
  title = {Rapid High-Fidelity Single-Shot Dispersive Readout of Superconducting Qubits},
  author = {Walter, T. and Kurpiers, P. and Gasparinetti, S. and Magnard, P. and Poto\ifmmode \check{c}\else \v{c}\fi{}nik, A. and Salath\'e, Y. and Pechal, M. and Mondal, M. and Oppliger, M. and Eichler, C. and Wallraff, A.},
  journal = {Phys. Rev. Appl.},
  volume = {7},
  issue = {5},
  pages = {054020},
  numpages = {11},
  year = {2017},
  month = {May},
  publisher = {American Physical Society},
  doi = {10.1103/PhysRevApplied.7.054020},
  url = {https://link.aps.org/doi/10.1103/PhysRevApplied.7.054020}
}

@article{Lienhard2022Deep,
  title = {Deep-Neural-Network Discrimination of Multiplexed Superconducting-Qubit States},
  author = {Lienhard, Benjamin and Veps\"al\"ainen, Antti and Govia, Luke C.G. and Hoffer, Cole R. and Qiu, Jack Y. and Rist\`e, Diego and Ware, Matthew and Kim, David and Winik, Roni and Melville, Alexander and Niedzielski, Bethany and Yoder, Jonilyn and Ribeill, Guilhem J. and Ohki, Thomas A. and Krovi, Hari K. and Orlando, Terry P. and Gustavsson, Simon and Oliver, William D.},
  journal = {Phys. Rev. Appl.},
  volume = {17},
  issue = {1},
  pages = {014024},
  numpages = {18},
  year = {2022},
  month = {Jan},
  publisher = {American Physical Society},
  doi = {10.1103/PhysRevApplied.17.014024},
  url = {https://link.aps.org/doi/10.1103/PhysRevApplied.17.014024}
}

@INPROCEEDINGS{Patel2020Experimental,

  author={Patel, Tirthak and Potharaju, Abhay and Li, Baolin and Roy, Rohan Basu and Tiwari, Devesh},

  booktitle={SC20: International Conference for High Performance Computing, Networking, Storage and Analysis}, 

  title={Experimental Evaluation of NISQ Quantum Computers: Error Measurement, Characterization, and Implications}, 

  year={2020},

  volume={},

  number={},

  pages={1-15},

  doi={10.1109/SC41405.2020.00050},

  ISSN={},

  month={Nov},}

@article{Macklin2015Near,
author = {C. Macklin  and K. O’Brien  and D. Hover  and M. E. Schwartz  and V. Bolkhovsky  and X. Zhang  and W. D. Oliver  and I. Siddiqi },
title = {A near–quantum-limited Josephson traveling-wave parametric amplifier},
journal = {Science},
volume = {350},
number = {6258},
pages = {307-310},
year = {2015},
doi = {10.1126/science.aaa8525},
URL = {https://www.science.org/doi/abs/10.1126/science.aaa8525},
eprint = {https://www.science.org/doi/pdf/10.1126/science.aaa8525}}

@article{Mallet2009Single,
  title={Single-shot qubit readout in circuit quantum electrodynamics},
  author={Mallet, Fran{\c{c}}ois and Ong, Florian R and Palacios-Laloy, Agustin and Nguyen, Francois and Bertet, Patrice and Vion, Denis and Esteve, Daniel},
  journal={Nature Physics},
  volume={5},
  number={11},
  pages={791--795},
  year={2009},
  publisher={Nature Publishing Group UK London}
}

@article{Thorbeck2024Readout,
  title = {Readout-Induced Suppression and Enhancement of Superconducting Qubit Lifetimes},
  author = {Thorbeck, Ted and Xiao, Zhihao and Kamal, Archana and Govia, Luke C. G.},
  journal = {Phys. Rev. Lett.},
  volume = {132},
  issue = {9},
  pages = {090602},
  numpages = {6},
  year = {2024},
  month = {Feb},
  publisher = {American Physical Society},
  doi = {10.1103/PhysRevLett.132.090602},
  url = {https://link.aps.org/doi/10.1103/PhysRevLett.132.090602}
}

@article{Chen2012Multiplexed,
    author = {Chen, Yu and Sank, D. and O'Malley, P. and White, T. and Barends, R. and Chiaro, B. and Kelly, J. and Lucero, E. and Mariantoni, M. and Megrant, A. and Neill, C. and Vainsencher, A. and Wenner, J. and Yin, Y. and Cleland, A. N. and Martinis, John M.},
    title = {Multiplexed dispersive readout of superconducting phase qubits},
    journal = {Applied Physics Letters},
    volume = {101},
    number = {18},
    pages = {182601},
    year = {2012},
    month = {11},
    issn = {0003-6951},
    doi = {10.1063/1.4764940},
    url = {https://doi.org/10.1063/1.4764940},
    eprint = {https://pubs.aip.org/aip/apl/article-pdf/doi/10.1063/1.4764940/14258287/182601\_1\_online.pdf},
}

@article{Motta2023Quantum,
  title={Quantum chemistry simulation of ground-and excited-state properties of the sulfonium cation on a superconducting quantum processor},
  author={Motta, Mario and Jones, Gavin O and Rice, Julia E and Gujarati, Tanvi P and Sakuma, Rei and Liepuoniute, Ieva and Garcia, Jeannette M and Ohnishi, Yu-ya},
  journal={Chemical Science},
  volume={14},
  number={11},
  pages={2915--2927},
  year={2023},
  publisher={Royal Society of Chemistry}
}

@article{Gidney2021How,
   title={How to factor 2048 bit RSA integers in 8 hours using 20 million noisy qubits},
   volume={5},
   ISSN={2521-327X},
   url={http://dx.doi.org/10.22331/q-2021-04-15-433},
   DOI={10.22331/q-2021-04-15-433},
   journal={Quantum},
   publisher={Verein zur Forderung des Open Access Publizierens in den Quantenwissenschaften},
   author={Gidney, Craig and Ekerå, Martin},
   year={2021},
   month=apr, pages={433} }

@article{Fowler2012Surface,
  title = {Surface codes: Towards practical large-scale quantum computation},
  author = {Fowler, Austin G. and Mariantoni, Matteo and Martinis, John M. and Cleland, Andrew N.},
  journal = {Phys. Rev. A},
  volume = {86},
  issue = {3},
  pages = {032324},
  numpages = {48},
  year = {2012},
  month = {Sep},
  publisher = {American Physical Society},
  doi = {10.1103/PhysRevA.86.032324},
  url = {https://link.aps.org/doi/10.1103/PhysRevA.86.032324}
}

@article{Bravyi2022Future,
    author = {Bravyi, Sergey and Dial, Oliver and Gambetta, Jay M. and Gil, Darío and Nazario, Zaira},
    title = {The future of quantum computing with superconducting qubits},
    journal = {Journal of Applied Physics},
    volume = {132},
    number = {16},
    pages = {160902},
    year = {2022},
    month = {10},
    issn = {0021-8979},
    doi = {10.1063/5.0082975},
    url = {https://doi.org/10.1063/5.0082975},
    eprint = {https://pubs.aip.org/aip/jap/article-pdf/doi/10.1063/5.0082975/20034201/160902\_1\_5.0082975.pdf},
}

@book{Nielsen2010Quantum, 
    place={Cambridge}, 
    title={Quantum Computation and Quantum Information: 10th Anniversary Edition}, 
    publisher={Cambridge University Press}, 
    author={Nielsen, Michael A. and Chuang, Isaac L.}, 
    year={2010}
}

@article{Shor1995Scheme,
  title = {Scheme for reducing decoherence in quantum computer memory},
  author = {Shor, Peter W.},
  journal = {Phys. Rev. A},
  volume = {52},
  issue = {4},
  pages = {R2493--R2496},
  numpages = {0},
  year = {1995},
  month = {Oct},
  publisher = {American Physical Society},
  doi = {10.1103/PhysRevA.52.R2493},
  url = {https://link.aps.org/doi/10.1103/PhysRevA.52.R2493}
}

@misc{Harper2025Characterising,
      title={Characterising the failure mechanisms of error-corrected quantum logic gates}, 
      author={Robin Harper and Constance Lainé and Evan Hockings and Campbell McLauchlan and Georgia M. Nixon and Benjamin J. Brown and Stephen D. Bartlett},
      year={2025},
      eprint={2504.07258},
      archivePrefix={arXiv},
      primaryClass={quant-ph},
      url={https://arxiv.org/abs/2504.07258}, 
}

@article{Battistel2023Real,
   title={Real-time decoding for fault-tolerant quantum computing: progress, challenges and outlook},
   volume={7},
   ISSN={2399-1984},
   url={http://dx.doi.org/10.1088/2399-1984/aceba6},
   DOI={10.1088/2399-1984/aceba6},
   number={3},
   journal={Nano Futures},
   publisher={IOP Publishing},
   author={Battistel, F and Chamberland, C and Johar, K and Overwater, R W J and Sebastiano, F and Skoric, L and Ueno, Y and Usman, M},
   year={2023},
   month=aug, pages={032003} }

@article{Gidney2021Stim,
  doi = {10.22331/q-2021-07-06-497},
  url = {https://doi.org/10.22331/q-2021-07-06-497},
  title = {Stim: a fast stabilizer circuit simulator},
  author = {Gidney, Craig},
  journal = {{Quantum}},
  issn = {2521-327X},
  publisher = {{Verein zur F{\"{o}}rderung des Open Access Publizierens in den Quantenwissenschaften}},
  volume = {5},
  pages = {497},
  month = jul,
  year = {2021}
}

@inproceedings{Maurya2025Synchronization, series={SIGARCH ’25},
   title={Synchronization for Fault-Tolerant Quantum Computers},
   url={http://dx.doi.org/10.1145/3695053.3730991},
   DOI={10.1145/3695053.3730991},
   booktitle={Proceedings of the 52nd Annual International Symposium on Computer Architecture},
   publisher={ACM},
   author={Maurya, Satvik and Tannu, Swamit},
   year={2025},
   month=jun, pages={1370–1385},
   collection={SIGARCH ’25} }

@inproceedings{Giortamis2025Mcmit,
      title={MCMit: Mid-Circuit Measurement Error Mitigation},
      author={Giortamis, Emmanouil and Gust, Felix and \'Swierkowska, Aleksandra and Stankovic, Sandra and Fulginiti, Innocenzo and Chen, Yanbin and Guo, Xiaorang and Lienhard, Benjamin and Schulz, Martin and Bhatotia, Pramod},
      year={2026},
      booktitle={Proceedings of the 59th IEEE/ACM International Symposium on Microarchitecture},
      series={MICRO '26},
      note={To appear},
      eprint={2604.25863},
      archivePrefix={arXiv},
      primaryClass={quant-ph},
      url={https://arxiv.org/abs/2604.25863},
}

@misc{Fruitwala2024Distributed,
      title={Distributed Architecture for FPGA-based Superconducting Qubit Control}, 
      author={Neelay Fruitwala and Gang Huang and Yilun Xu and Abhi Rajagopala and Akel Hashim and Ravi K. Naik and Kasra Nowrouzi and David I. Santiago and Irfan Siddiqi},
      year={2024},
      eprint={2404.15260},
      archivePrefix={arXiv},
      primaryClass={quant-ph},
      url={https://arxiv.org/abs/2404.15260}, 
}

@INPROCEEDINGS {Chatterjee2024Magic,
author = { Chatterjee, Avimita and Ghosh, Swaroop },
booktitle = { 2024 IEEE International Conference on Quantum Computing and Engineering (QCE) },
title = {{ Magic Mirror on the Wall, How to Benchmark Quantum Error Correction Codes, Overall? }},
year = {2024},
volume = {},
ISSN = {},
pages = {356-367},
doi = {10.1109/QCE60285.2024.00050},
url = {https://doi.ieeecomputersociety.org/10.1109/QCE60285.2024.00050},
publisher = {IEEE Computer Society},
address = {Los Alamitos, CA, USA},
month =sep}

@article{Bravyi2024High,
author={Bravyi, Sergey
and Cross, Andrew W.
and Gambetta, Jay M.
and Maslov, Dmitri
and Rall, Patrick
and Yoder, Theodore J.},
title={High-threshold and low-overhead fault-tolerant quantum memory},
journal={Nature},
year={2024},
month={Mar},
day={01},
volume={627},
number={8005},
pages={778-782},
issn={1476-4687},
doi={10.1038/s41586-024-07107-7},
url={https://doi.org/10.1038/s41586-024-07107-7}
}

@article{Breuckmann2021Quantum,
  title = {Quantum Low-Density Parity-Check Codes},
  author = {Breuckmann, Nikolas P. and Eberhardt, Jens Niklas},
  journal = {PRX Quantum},
  volume = {2},
  issue = {4},
  pages = {040101},
  numpages = {19},
  year = {2021},
  month = {Oct},
  publisher = {American Physical Society},
  doi = {10.1103/PRXQuantum.2.040101},
  url = {https://link.aps.org/doi/10.1103/PRXQuantum.2.040101}
}

@article{Steane1996Error,
  title = {Error Correcting Codes in Quantum Theory},
  author = {Steane, A. M.},
  journal = {Phys. Rev. Lett.},
  volume = {77},
  issue = {5},
  pages = {793--797},
  numpages = {0},
  year = {1996},
  month = {Jul},
  publisher = {American Physical Society},
  doi = {10.1103/PhysRevLett.77.793},
  url = {https://link.aps.org/doi/10.1103/PhysRevLett.77.793}
}

@article{Katsuda2024Simulation,
  title = {Simulation and performance analysis of quantum error correction with a rotated surface code under a realistic noise model},
  author = {Katsuda, Mitsuki and Mitarai, Kosuke and Fujii, Keisuke},
  journal = {Phys. Rev. Res.},
  volume = {6},
  issue = {1},
  pages = {013024},
  numpages = {9},
  year = {2024},
  month = {Jan},
  publisher = {American Physical Society},
  doi = {10.1103/PhysRevResearch.6.013024},
  url = {https://link.aps.org/doi/10.1103/PhysRevResearch.6.013024}
}

@article{Wootton2020Benchmarking,
doi = {10.1088/2058-9565/aba038},
url = {https://dx.doi.org/10.1088/2058-9565/aba038},
year = {2020},
month = {jul},
publisher = {IOP Publishing},
volume = {5},
number = {4},
pages = {044004},
author = {Wootton, James R},
title = {Benchmarking near-term devices with quantum error correction},
journal = {Quantum Science and Technology}
}

@article{Lorenz2025Systematic,
    author = "Lorenz, Jeanette Miriam and others",
    title = "{Systematic benchmarking of quantum computers: status and recommendations}",
    eprint = "2503.04905",
    archivePrefix = "arXiv",
    primaryClass = "quant-ph",
    month = "3",
    year = "2025"
}

@INPROCEEDINGS{Tomesh2022Supermarq,
author = { Tomesh, Teague and Gokhale, Pranav and Omole, Victory and Ravi, Gokul Subramanian and Smith, Kaitlin N. and Viszlai, Joshua and Wu, Xin-Chuan and Hardavellas, Nikos and Martonosi, Margaret R. and Chong, Frederic T. },
booktitle = { 2022 IEEE International Symposium on High-Performance Computer Architecture (HPCA) },
title = {{ SupermarQ: A Scalable Quantum Benchmark Suite }},
year = {2022},
volume = {},
ISSN = {},
pages = {587-603},
doi = {10.1109/HPCA53966.2022.00050},
url = {https://doi.ieeecomputersociety.org/10.1109/HPCA53966.2022.00050},
publisher = {IEEE Computer Society},
address = {Los Alamitos, CA, USA},
month =apr}

@article{Li2022Qasmbench,
author = {Li, Ang and Stein, Samuel and Krishnamoorthy, Sriram and Ang, James},
year = {2022},
month = {07},
pages = {},
title = {QASMBench: A Low-Level Quantum Benchmark Suite for NISQ Evaluation and Simulation},
volume = {4},
journal = {ACM Transactions on Quantum Computing},
doi = {10.1145/3550488}
}

@article{Finsterhoelzl2022Benchmarking,
doi = {10.1088/2058-9565/aca21f},
url = {https://dx.doi.org/10.1088/2058-9565/aca21f},
year = {2022},
month = {nov},
publisher = {IOP Publishing},
volume = {8},
number = {1},
pages = {015013},
author = {Finsterhoelzl, Regina and Burkard, Guido},
title = {Benchmarking quantum error-correcting codes on quasi-linear and central-spin processors},
journal = {Quantum Science and Technology}
}

@article{Nation2025Benchmarking,
author = {Nation, Paul and Saki, Abdullah Ash and Brandhofer, Sebastian and Bello, Luciano and Garion, Shelly and Treinish, Matthew and Javadi-Abhari, Ali},
year = {2025},
month = {04},
pages = {1-9},
title = {Benchmarking the performance of quantum computing software for quantum circuit creation, manipulation and compilation},
journal = {Nature Computational Science},
doi = {10.1038/s43588-025-00792-y}
}

@misc{Mckay2023Benchmarking,
      title={Benchmarking Quantum Processor Performance at Scale}, 
      author={David C. McKay and Ian Hincks and Emily J. Pritchett and Malcolm Carroll and Luke C. G. Govia and Seth T. Merkel},
      year={2023},
      eprint={2311.05933},
      archivePrefix={arXiv},
      primaryClass={quant-ph},
      url={https://arxiv.org/abs/2311.05933}, 
}

@misc{Proctor2024Benchmarking,
      title={Benchmarking quantum computers}, 
      author={Timothy Proctor and Kevin Young and Andrew D. Baczewski and Robin Blume-Kohout},
      year={2024},
      eprint={2407.08828},
      archivePrefix={arXiv},
      primaryClass={quant-ph},
      url={https://arxiv.org/abs/2407.08828}, 
}

@misc{Kurman2024Benchmarking,
      title={Benchmarking the ability of a controller to execute quantum error corrected non-Clifford circuits}, 
      author={Yaniv Kurman and Lior Ella and Ramon Szmuk and Oded Wertheim and Benedikt Dorschner and Sam Stanwyck and Yonatan Cohen},
      year={2024},
      eprint={2311.07121},
      archivePrefix={arXiv},
      primaryClass={quant-ph},
      url={https://arxiv.org/abs/2311.07121}, 
}

@misc{Zhao2024Benchmarking,
      title={Benchmarking Machine Learning Models for Quantum Error Correction}, 
      author={Yue Zhao},
      year={2024},
      eprint={2311.11167},
      archivePrefix={arXiv},
      primaryClass={quant-ph},
      url={https://arxiv.org/abs/2311.11167}, 
}

@article{Tomita2014Low,
   title={Low-distance surface codes under realistic quantum noise},
   volume={90},
   ISSN={1094-1622},
   url={http://dx.doi.org/10.1103/PhysRevA.90.062320},
   DOI={10.1103/physreva.90.062320},
   number={6},
   journal={Physical Review A},
   publisher={American Physical Society (APS)},
   author={Tomita, Yu and Svore, Krysta M.},
   year={2014},
   month=dec }

@Article{Acharya2025Quantum,
author={{Google Quantum AI and Collaborators}},
title={Quantum error correction below the surface code threshold},
journal={Nature},
year={2025},
month={Feb},
day={01},
volume={638},
number={8052},
pages={920-926},
issn={1476-4687},
doi={10.1038/s41586-024-08449-y},
url={https://doi.org/10.1038/s41586-024-08449-y}
}

@article{Bacon2006Operator,
   title={Operator quantum error-correcting subsystems for self-correcting quantum memories},
   volume={73},
   ISSN={1094-1622},
   url={http://dx.doi.org/10.1103/PhysRevA.73.012340},
   DOI={10.1103/physreva.73.012340},
   number={1},
   journal={Physical Review A},
   publisher={American Physical Society (APS)},
   author={Bacon, Dave},
   year={2006},
   month=jan }

@article{Wootters1982Single,
    author = "Wootters, W. K. and Zurek, W. H.",
    title = "{A single quantum cannot be cloned}",
    doi = "10.1038/299802a0",
    journal = "Nature",
    volume = "299",
    pages = "802--803",
    year = "1982"
}

@article{Krantz2019Quantum,
   title={A quantum engineer’s guide to superconducting qubits},
   volume={6},
   ISSN={1931-9401},
   url={http://dx.doi.org/10.1063/1.5089550},
   DOI={10.1063/1.5089550},
   number={2},
   journal={Applied Physics Reviews},
   publisher={AIP Publishing},
   author={Krantz, P. and Kjaergaard, M. and Yan, F. and Orlando, T. P. and Gustavsson, S. and Oliver, W. D.},
   year={2019},
   month=jun }

@article{Roffe2020Decoding,
   title={Decoding across the quantum low-density parity-check code landscape},
   volume={2},
   ISSN={2643-1564},
   url={http://dx.doi.org/10.1103/PhysRevResearch.2.043423},
   DOI={10.1103/physrevresearch.2.043423},
   number={4},
   journal={Physical Review Research},
   publisher={American Physical Society (APS)},
   author={Roffe, Joschka and White, David R. and Burton, Simon and Campbell, Earl},
   year={2020},
   month=dec }

@article{Chamberland2020Topological,
   title={Topological and Subsystem Codes on Low-Degree Graphs with Flag Qubits},
   volume={10},
   ISSN={2160-3308},
   url={http://dx.doi.org/10.1103/PhysRevX.10.011022},
   DOI={10.1103/physrevx.10.011022},
   number={1},
   journal={Physical Review X},
   publisher={American Physical Society (APS)},
   author={Chamberland, Christopher and Zhu, Guanyu and Yoder, Theodore J. and Hertzberg, Jared B. and Cross, Andrew W.},
   year={2020},
   month=jan }

@article{Chamberland2020Triangular,
   title={Triangular color codes on trivalent graphs with flag qubits},
   volume={22},
   ISSN={1367-2630},
   url={http://dx.doi.org/10.1088/1367-2630/ab68fd},
   DOI={10.1088/1367-2630/ab68fd},
   number={2},
   journal={New Journal of Physics},
   publisher={IOP Publishing},
   author={Chamberland, Christopher and Kubica, Aleksander and Yoder, Theodore J and Zhu, Guanyu},
   year={2020},
   month=feb, pages={023019} }

@misc{Knill1996Concatenated,
      title={Concatenated Quantum Codes}, 
      author={Emanuel Knill and Raymond Laflamme},
      year={1996},
      eprint={quant-ph/9608012},
      archivePrefix={arXiv},
      primaryClass={quant-ph},
      url={https://arxiv.org/abs/quant-ph/9608012}, 
}

@article{Bombin2006Topological,
   title={Topological Quantum Distillation},
   volume={97},
   ISSN={1079-7114},
   url={http://dx.doi.org/10.1103/PhysRevLett.97.180501},
   DOI={10.1103/physrevlett.97.180501},
   number={18},
   journal={Physical Review Letters},
   publisher={American Physical Society (APS)},
   author={Bombin, H. and Martin-Delgado, M. A.},
   year={2006},
   month=oct }

@INPROCEEDINGS{Shor1994Algorithms,
  author={Shor, P.W.},
  booktitle={Proceedings 35th Annual Symposium on Foundations of Computer Science}, 
  title={Algorithms for quantum computation: discrete logarithms and factoring}, 
  year={1994},
  volume={},
  number={},
  pages={124-134},
  doi={10.1109/SFCS.1994.365700}}

@article{Dennis2002Topological,
   title={Topological quantum memory},
   volume={43},
   ISSN={1089-7658},
   url={http://dx.doi.org/10.1063/1.1499754},
   DOI={10.1063/1.1499754},
   number={9},
   journal={Journal of Mathematical Physics},
   publisher={AIP Publishing},
   author={Dennis, Eric and Kitaev, Alexei and Landahl, Andrew and Preskill, John},
   year={2002},
   month=sep, pages={4452–4505} }

@article{Foumani2024,
author = {Mohammadi Foumani, Navid and Miller, Lynn and Tan, Chang Wei and Webb, Geoffrey I. and Forestier, Germain and Salehi, Mahsa},
title = {Deep Learning for Time Series Classification and Extrinsic Regression: A Current Survey},
year = {2024},
issue_date = {September 2024},
publisher = {Association for Computing Machinery},
address = {New York, NY, USA},
volume = {56},
number = {9},
issn = {0360-0300},
url = {https://doi.org/10.1145/3649448},
doi = {10.1145/3649448},
journal = {ACM Comput. Surv.},
month = apr,
articleno = {217},
numpages = {45}
}

@article{Harper2026Characterising,
  title = {Characterising the failure mechanisms of error-corrected quantum logic gates},
  author = {Harper, Robin and Lain{\'e}, Constance and Hockings, Evan and McLauchlan, Campbell and Nixon, Georgia M. and Brown, Benjamin J. and Bartlett, Stephen D.},
  journal = {Nature Communications},
  volume = {17},
  pages = {5039},
  year = {2026},
  publisher = {Nature Publishing Group},
  doi = {10.1038/s41467-026-71773-6},
  url = {https://www.nature.com/articles/s41467-026-71773-6}
}

\newpage
\appendix

\section{Qubit-State Readout in the Dispersive Regime}
\label{app:readout}

This appendix expands on qubit-state readout in the dispersive regime summarized in
\S\ref{subsec:readout}.

\myparagraph{Measurements in the dispersive regime} In state-of-the-art superconducting architectures, a qubit is
not measured directly but through a microwave resonator coupled to it. In the
dispersive regime (where the qubit-resonator detuning is large compared to their
coupling), this interaction shifts the resonator frequency conditional
on the qubit state, without exchanging energy with the qubit~\cite{Mallet2009Single,
Krantz2019Quantum}. Probing the resonator with a microwave tone, therefore, returns a
signal whose phase and amplitude carry the state information, while the weak coupling
keeps the qubit largely undisturbed~\cite{Walter2017Rapid}. This is a deliberate
trade-off: weak coupling protects the extracted quantum state during measurement, but yields a
faint signal, so the returned tone must be boosted by a near-quantum-limited
amplifier and integrated over time to be discriminated reliably~\cite{Macklin2015Near,
Chen2023Transmon}.

\myparagraph{From signal to outcome} The amplified signal is analog demodulated, digitized, 
and accumulated into a stream of In-Phase (I) and Quadrature (Q) samples. Repeated
single-shot measurements of a prepared $|0\rangle$ or $|1\rangle$ populate two clouds
in the IQ plane; their separation relative to their spread sets the achievable
readout fidelity~\cite{Heinsoo2018Rapid,Magesan2015Machine}. State discrimination
reduces to classifying which cloud a trace belongs to, traditionally with a static
linear boundary and increasingly with learned discriminators that capture the
nonlinear, time-dependent structure of the trace~\cite{Lienhard2022Deep,
Maurya2023Scaling}.

\myparagraph{Practical error sources} Beyond the inherent speed-fidelity trade-off,
two effects degrade readout in practice. \emph{Measurement-induced dephasing} and
readout-induced state transitions perturb the qubit while it is being measured,
coupling readout strength to qubit lifetime~\cite{Thorbeck2024Readout}. \emph{Crosstalk}
arises in frequency-multiplexed architectures, where many qubit resonators share a
single feedline: neighboring tones leak into one another and blur the IQ clusters,
making discrimination more ambiguous~\cite{Tannu2019Mitigating,Heinsoo2018Rapid}.
These effects are difficult to model analytically, which is precisely why
data-driven discriminators (\S\ref{subsec:ml-discrimination}) are attractive.

\section{QPU Error Rates Across Vendors}
\label{app:error-rates}

Table~\ref{tab:readout-errors} reports median error rates on QPUs from four major vendors, showing that the readout error exceeds single- and two-qubit gate errors by up to an order of magnitude across the industry. Absolute rates fluctuate with calibration; the relative ordering is consistent.

\begin{table}[h]
\centering
\footnotesize
\setlength{\tabcolsep}{5pt}
\caption{Median error rates on QPUs from four major vendors. The readout error exceeds single- and two-qubit gate errors by up to an order of magnitude. Absolute rates fluctuate with calibration.}
\label{tab:readout-errors}
\begin{tabular}{lccc}
\toprule
QPU & Single-qubit gate & Two-qubit gate & Readout \\
\midrule
IBM Heron~\cite{UnknownXXXXIbma}        & 0.03\% & 0.3\% & 1.3\% \\
Google Willow~\cite{Acharya2025Quantum} & 0.04\% & 0.3\% & 0.7\% \\
IQM Garnet~\cite{IQM2024Garnet}         & 0.05\% & 0.5\% & 2.0\% \\
Rigetti Ankaa-3~\cite{Rigetti2024Ankaa} & 0.10\% & 1.0\% & 4.0\% \\
\bottomrule
\end{tabular}
\end{table}

\section{HERQULES: Evaluation Protocol}
\label{app:herqules-bug}

\textsc{HERQULES}~\cite{Maurya2023Scaling} classifies a five-qubit readout by
applying a matched filter (MF) to each demodulated trace and feeding the MF and
relaxation-MF scores to a small ($\sim$1000-parameter) network. We evaluate it
on the full dataset; the seed-to-seed variance is $\approx0.0002$, so
the figures below are deterministic and reproducible rather than a matter of luck.
For transparency, this appendix documents two protocol choices where our
deployment-faithful evaluation departs from the released reference pipeline, and
what each is worth in fidelity.

\myparagraph{Feature-vector assembly}
We read out all qubits simultaneously, so a single shot yields one MF score per
qubit, and the deployment-realistic feature vector concatenates the five scores
\emph{from the same} shot. The released feature-assembly routine instead builds the
vector one qubit (column) at a time and independently re-orders the shots within
each basis-state block, so a row can mix scores from different physical
acquisitions; the labels are unaffected (all shots in a block share the same
state), but inter-qubit correlations (crosstalk, common-mode noise, coincident
errors) are averaged out. Throughout this paper, we instead draw each feature vector
from a single shot, matching how a real-time controller would assemble it. On this
dataset, the difference is small (per-qubit fidelities are unchanged and the joint
fidelity moves only marginally), so it is not the source of the gap discussed next.

A second, shot-aligned choice remains even once feature vectors are drawn from
single shots: whether the MF table is built directly from shots in acquisition
order (\emph{per-trace}) or from a per-state random subsample, as in HERQULES's
released training code (\emph{demux-subsample}). Both are valid, shot-aligned
constructions; they differ by a margin that looks small in raw fidelity but is
larger once expressed as error reduction (Table~\ref{tab:herqules-pertrace}):
\begin{itemize}
  \item \textbf{Per-trace ($F_{5Q}=0.905$), deployment-faithful.} Every feature row
  is computed from one physical shot and used as-is, matching real-time
  deployment: a shot arrives, its MF scores are computed, and it is classified.
  \item \textbf{Demux-subsample ($F_{5Q}=0.925$), reference training recipe.} The MF
  step in HERQULES's own training code randomly subsamples and reshuffles the shots
  \emph{per prepared state} before building the table, yielding a table the small
  network fits somewhat more easily.
\end{itemize}
Under the $(\text{better}-\text{worse})/(100-\text{worse})$ error-reduction metric,
the demux recipe removes $\sim$21\% of the per-trace error at $1\,\mu$s, and the
gap widens at shorter windows, to $\sim$31--35\% below $600$\,ns ($F_{5Q}=0.794$
demux vs.\ $0.702$ per-trace at $200$\,ns). We report per-trace numbers throughout
the main text (Table~\ref{tab:fidelity}) since it reflects real-time deployment, particularly at the
short readout windows our analysis targets
(\S\ref{subsec:discriminator_perforamnce}); we include the demux numbers here for
comparability with previously reported figures.

\begin{table}[h]
\centering
\caption{HERQULES under our deployment-faithful \emph{per-trace} protocol (every
feature row computed from one physical shot) versus the \emph{demux-subsample}
assembly of the reference training pipeline (shot-aligned, but resampled per
prepared state).}
\label{tab:herqules-pertrace}
\begin{tabular}{l cc cc}
\toprule
& \multicolumn{2}{c}{Per-trace (ours)} & \multicolumn{2}{c}{As released} \\
\cmidrule(lr){2-3}\cmidrule(lr){4-5}
Duration & $F_{5Q}$ & $F_{4Q}$ & $F_{5Q}$ & $F_{4Q}$ \\
\midrule
1000\,ns & 0.905 & 0.951 & 0.925 & 0.973 \\
800\,ns  & 0.901 & 0.949 & 0.926 & 0.973 \\
600\,ns  & 0.894 & 0.943 & 0.927 & 0.972 \\
400\,ns  & 0.867 & 0.917 & 0.914 & 0.961 \\
200\,ns  & 0.702 & 0.759 & 0.794 & 0.825 \\
\bottomrule
\end{tabular}
\end{table}

\myparagraph{Retraining across durations}
The original HERQULES work~\cite{Maurya2023Scaling} additionally reports that its
matched-filter preprocessing supports classifying shorter traces with a model
trained on the full-duration trace, without retraining. Throughout the main text
(Table~\ref{tab:fidelity}) we instead
retrain HERQULES, like every other discriminator, from scratch at each readout
window, so that all models receive the same training budget at every duration. For
completeness, Table~\ref{tab:herqules-noretrain} reports HERQULES under its own
no-retraining protocol, i.e.\ a single model trained on the full $1\,\mu$s trace and
evaluated directly on truncated inputs.

Under this no-retraining protocol, fidelity degrades as the window shrinks: the
$1\,\mu$s row reproduces the retrained, per-trace HERQULES numbers exactly
($F_{5Q}=0.905$), falling to $0.881$ at $800$\,ns, $0.775$ at $600$\,ns, $0.628$ at
$400$\,ns, and $0.504$ at $200$\,ns (chance level on every qubit). Retraining at
each duration, our protocol throughout the main text, instead yields
$0.894 / 0.867 / 0.702$ at $600 / 400 / 200$\,ns (Table~\ref{tab:fidelity}). On this
dataset, then, retraining at each duration is needed to reach the fidelities we
report, which is why we adopt it uniformly across all discriminators.

\begin{table}[h]
\centering
\caption{HERQULES evaluated \emph{without retraining}: a single model trained on
the full $1\,\mu$s trace and applied directly to truncated inputs. $F_{5Q}$ is
the five-qubit geometric mean.}
\label{tab:herqules-noretrain}
\begin{tabular}{l rrrrr r}
\toprule
Duration & Qubit 1 & Qubit 2 & Qubit 3 & Qubit 4 & Qubit 5 & $F_{5Q}$ \\
\midrule
1000\,ns & 0.969 & 0.740 & 0.924 & 0.944 & 0.970 & 0.905 \\
800\,ns  & 0.887 & 0.727 & 0.923 & 0.921 & 0.970 & 0.881 \\
600\,ns  & 0.590 & 0.692 & 0.920 & 0.768 & 0.965 & 0.775 \\
400\,ns  & 0.500 & 0.624 & 0.847 & 0.512 & 0.722 & 0.628 \\
200\,ns  & 0.500 & 0.522 & 0.500 & 0.500 & 0.500 & 0.504 \\
\bottomrule
\end{tabular}
\end{table}

\section{Integration-and-Threshold Baseline Fidelity}
\label{app:linear-baseline}

Table~\ref{tab:fidelity_linear} reports per-qubit fidelity vs.\ trace duration for the
integration-and-threshold baseline (the \emph{Linear Threshold} discriminator of
\S\ref{sec:methodology}), omitted from Table~\ref{tab:fidelity} in the main text. It performs similar to QubiCML and attains some per-qubit
optima below the full $1\,\mu$s window.

\begin{table}[h]
\centering
\small
\caption{Linear-threshold baseline: per-qubit readout fidelity vs.\ trace duration (cf.\ Table~\ref{tab:fidelity}). \textbf{Bold} = best duration for that qubit.}
\label{tab:fidelity_linear}
\begin{tabular}{l rrrrr rr}
\toprule
Duration & Qubit 1 & Qubit 2 & Qubit 3 & Qubit 4 & Qubit 5 & $F_{5Q}$ & $F_{4Q}$ \\
\midrule
1000\,ns & \textbf{0.967} & \textbf{0.730} & 0.894 & 0.936 & 0.958 & 0.892 & 0.938 \\
800\,ns & 0.961 & 0.726 & 0.903 & \textbf{0.936} & 0.936 & \textbf{0.893} & 0.934 \\
600\,ns & 0.947 & 0.715 & 0.910 & 0.930 & \textbf{0.965} & 0.888 & \textbf{0.938} \\
400\,ns & 0.900 & 0.693 & \textbf{0.910} & 0.898 & 0.942 & 0.864 & 0.912 \\
200\,ns & 0.753 & 0.635 & 0.840 & 0.749 & 0.731 & 0.739 & 0.767 \\
\bottomrule
\end{tabular}
\end{table}

\section{Error-Mode Taxonomy and Attribution}
\label{app:error-modes}

This appendix expands the failure analysis of \S\ref{subsec:failures}. In the integrated
IQ plane (Fig.~\ref{fig:iq_taxonomy}) each prepared computational state forms an
approximately Gaussian cloud, and the discriminator places a decision boundary between
the $|0\rangle$ and $|1\rangle$ clouds. A misclassified shot can arise from several
physically distinct mechanisms:

\begin{itemize}
\item \textbf{Near-boundary, low-SNR.} The integrated $(\bar I,\bar Q)$ point lands in the
overlap region between the two clouds. At any finite signal-to-noise ratio, the clouds
overlap, so a shot in the overlap falls on the wrong side through measurement noise
alone; no mid-trace event occurs, and no classifier, however expressive, can recover
the lost information. This is a property of the readout SNR, not of the model, and it
dominates the error budget in practice.
\item \textbf{Mid-readout $T_1$ decay (relaxation/excitation).} The qubit changes state
during the measurement (\emph{relaxation} ($|1\rangle\!\to\!|0\rangle$) or thermal
\emph{excitation} ($|0\rangle\!\to\!|1\rangle$)), producing a trace whose early and late
portions encode different states. The IQ trajectory shows a characteristic
blob-to-blob inflection, and the ``true'' label is itself ambiguous because it depends
on the unknown transition time. Longer windows expose more of the post-transition
trajectory, so this mode grows with measurement duration.
\item \textbf{Leakage / amplitude outlier.} Anomalous trace energy consistent with
population leaking outside the computational subspace, e.g.\ to the non-computational
$|2\rangle$ level, which lands in a displaced cloud of its own.
\item \textbf{Other.} Any misclassification not captured above, such as residual
state-preparation errors.
\end{itemize}

\myparagraph{Attribution procedure} We assign every misclassified shot to exactly one
mode. A shot is \emph{near-boundary} if its integrated $(\bar I,\bar Q)$ point lies in
the low-SNR overlap band between the two clouds; \emph{mid-readout $T_1$ decay} if the
trace exhibits a relaxation inflection (a blob-to-blob trajectory crossing) within the
window; \emph{leakage/amplitude outlier} if the trace energy is anomalous relative to
both computational clouds; and \emph{other} otherwise. We report per-mode rates for the
demux-pipeline HERQULES (the reference-training assembly of App.~\ref{app:herqules-bug},
which yields the fewest errors of any configuration we logged) and the strongest
deployable model MCMit-CNN, on a mid-SNR qubit (Q3) and the best qubit (Q5) at the full
$1\,\mu$s and truncated $400$\,ns windows (Fig.~\ref{fig:error_breakdown}), with the
full five-qubit MCMit-CNN breakdown at $1\,\mu$s in Table~\ref{tab:error_modes}.
\section{ECCentric vs.\ lattice-sim: Simulator Comparison}
\label{app:sim-comparison}
\label{app:futuristic}

\begin{figure}[h]
\centering
\includegraphics[width=\linewidth]{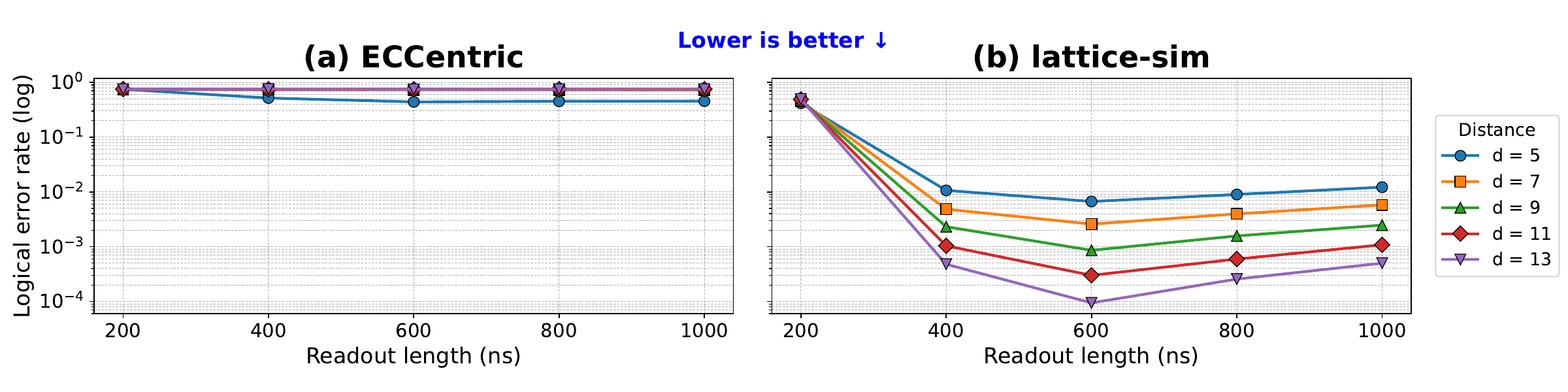}
\caption{LER (log scale) of a surface-code lattice-surgery CNOT vs.\ measurement duration under \textbf{current} IBM Heron~r3 noise: \textbf{(a)}~ECCentric, \textbf{(b)}~lattice-sim. \textit{ECCentric is above threshold (LER flat); lattice-sim is U-shaped and orders of magnitude lower, reflecting its narrower noise-model scope.}}
\label{fig:ler_current}
\end{figure}

\begin{figure}[h]
\centering
\includegraphics[width=\linewidth]{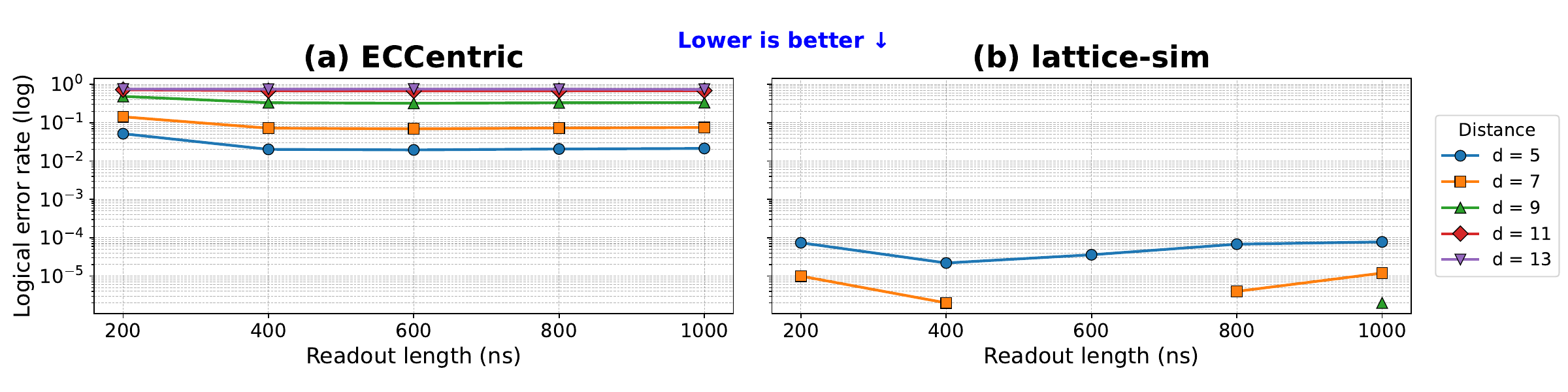}
\caption{Surface-code lattice-surgery CNOT under \textbf{futuristic} noise (all errors $/10$, $T_1/T_2$ tripled), ECCentric vs.\ lattice-sim (\emph{lower is better}). \textit{Same trend as current noise, shifted down: lattice-sim stays U-shaped and orders of magnitude lower; ECCentric plateaus beyond $\sim$600\,ns.}}
\label{fig:ler_futuristic}
\end{figure}

Figs.~\ref{fig:ler_current}--\ref{fig:ler_futuristic} show the surface-code lattice-surgery CNOT swept over measurement duration under current and futuristic noise in both simulators. The two tools differ in modeled noise scope: lattice-sim's released model covers $X$-type idle errors and a single logical observable, a narrower channel set than ECCentric's. Under current IBM Heron~r3 noise, this shows up as a clear divergence: ECCentric places the circuit above threshold (LER saturated near $0.5$, flat across measurement durations), while lattice-sim's narrower scope yields a pronounced U-shaped optimum near $600$\,ns at LERs orders of magnitude lower. The same pattern holds under futuristic noise (all errors $/10$, $T_1/T_2$ tripled): the measurement-duration trend is intact, just shifted down---ECCentric again plateaus beyond $\sim$400--600\,ns while lattice-sim remains U-shaped and orders of magnitude lower. The plateau also holds across all six codes in memory (\S\ref{subsec:regimes_qec}), ruling out a surface-code-specific result. Even under this projected noise, the LER still rises with code distance (the $/10$ scaling lowers the prefactor without crossing the threshold), so the measurement-duration benefit is a fixed-distance, low-LER effect rather than a fault-tolerance one. Given its broader noise-model coverage, we adopt ECCentric throughout for the discriminator comparisons in this paper.

\section{Decoherence-Only Noise Model}
\label{app:decoherence-only}

\begin{figure}[h]
\centering
\includegraphics[width=\linewidth]{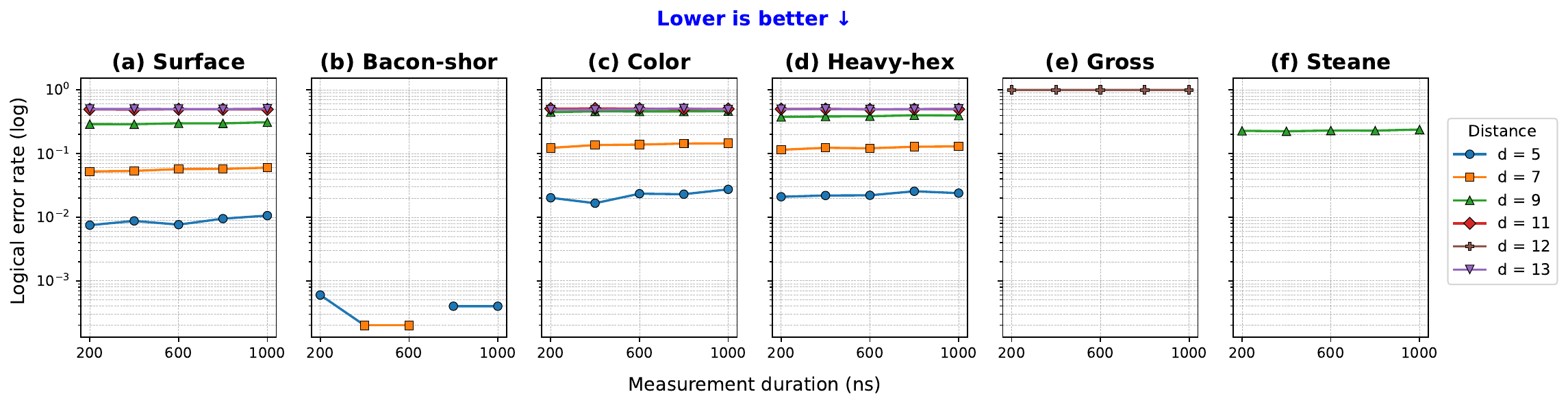}
\caption{LER vs.\ measurement duration, six codes, \textbf{decoherence-only} model (readout flip removed). \textit{All curves flatten, so the optimum is caused entirely by the flip error. Bacon-Shor falls to $\approx0$ (vs.\ $\approx0.5$ under the flip); Gross fails regardless.}}
\label{fig:six_decoh}
\end{figure}

Fig.~\ref{fig:six_decoh} repeats the six-code memory sweep of \S\ref{subsec:regimes_qec} with the readout flip removed entirely, leaving only idle decoherence. Every curve flattens: without a length-dependent flip error, the LER is essentially independent of measurement duration, confirming that the $\sim$600\,ns optimum and the short-trace catastrophe in Fig.~\ref{fig:six_current} are caused entirely by the flip error. The code dependence also sharpens: Bacon-Shor, whose syndromes rest on single-qubit measurements, collapses from $\approx0.5$ under the flip to $\approx0$ without it, while the gate-limited Gross code fails regardless.

\section{Fixed-Fidelity (Duration-Independent) Readout}
\label{app:fixedfidelity}

\begin{figure}[h]
\centering
\includegraphics[width=\linewidth]{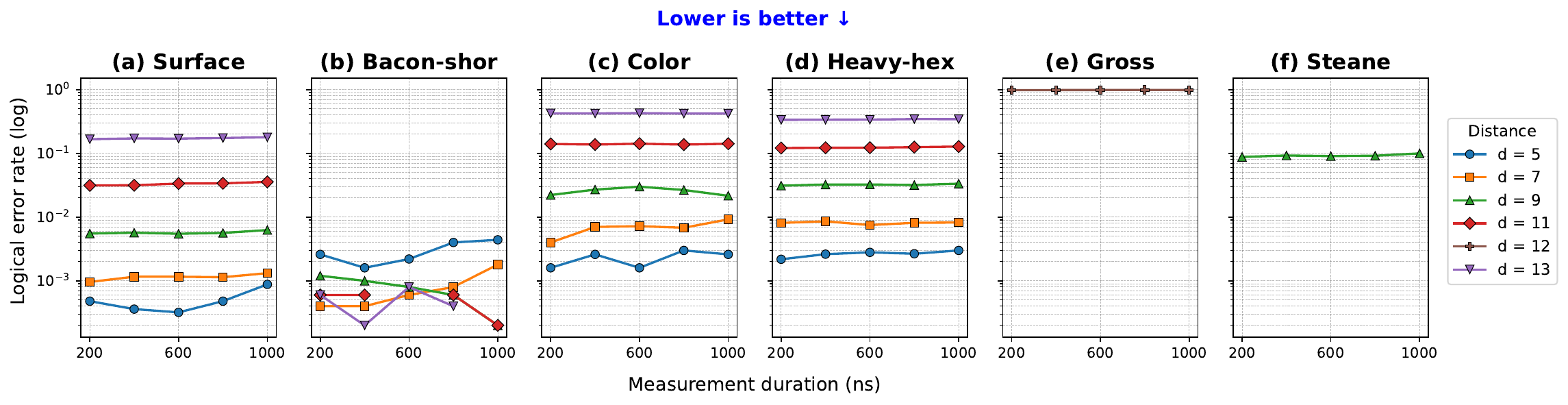}
\caption{LER vs.\ measurement duration, six codes, \textbf{fixed-fidelity} (duration-independent) readout on the futuristic sweep (ECCentric; \emph{lower is better}). \textit{The short-trace penalty disappears, and the LER falls monotonically as measurement shortens; a naive fixed-fidelity model would miss the optimum.}}
\label{fig:six_fut_fixed}
\end{figure}

Modeling readout as a \emph{constant}, duration-independent error (as a standard QEC
simulator would) removes the short-trace flip penalty entirely
(Fig.~\ref{fig:six_fut_fixed}): the LER then falls monotonically as the window
shortens, so a naive fixed-fidelity simulator would conclude ``always go shorter'' and
miss the $\sim$600\,ns optimum of \S\ref{subsec:regimes_qec}. This is precisely why we
derive a \emph{duration-dependent} per-shot flip error from real traces
(\S\ref{sec:methodology}) rather than plugging in a single calibration number; the
contrast (fixed- vs.\ duration-dependent readout, on the same futuristic sweep) is what
separates a faithful QEC simulator from a naive one.

\end{document}